%% file: main.tex
\documentclass[journal]{IEEEtran}

\usepackage[hidelinks]{hyperref} 

\usepackage{amssymb}
\usepackage{amsmath}
\usepackage{graphicx}
\usepackage{color}
\usepackage{multirow}
\usepackage{tabto}
\usepackage{float}
\usepackage[table,xcdraw,dvipsnames]{xcolor}
\usepackage{algpseudocode}
\usepackage{mathtools}

\usepackage{caption}
\usepackage{subcaption}

\usepackage[nopostdot,nogroupskip,nonumberlist]{glossaries}
\input{list_abbreviations}

\usepackage{cleveref}
\crefformat{section}{\S#2#1#3}
\crefformat{subsection}{\S#2#1#3}
\crefformat{subsubsection}{\S#2#1#3}

\usepackage{enumitem}
\setlist[itemize]{noitemsep, topsep=0pt, leftmargin=*}

\usepackage{courier}

\usepackage[T1]{fontenc}
\usepackage{comment}
\usepackage[ruled, vlined, linesnumbered]{algorithm2e}

\definecolor{blue_color}{rgb}{0.0, 0.0, 0.61}

\definecolor{red_color}{rgb}{0.59, 0.0, 0.09}
\SetKwProg{MyStruct}{struct}{}{end}
\SetKw{KwBy}{by}
\usepackage{cite}
\ifCLASSINFOpdf
\else
\fi

\begin{document}

\bstctlcite{IEEEexample:BSTcontrol}


\title{quicSDN: Transitioning from TCP to QUIC for Southbound Communication in SDNs}


\author{\IEEEauthorblockN{Puneet Kumar and Behnam Dezfouli}\\
\IEEEauthorblockA{\small Internet of Things Research Lab, Department of Computer Science and Engineering, Santa Clara University, USA
\\
\texttt{\small \{pkumar, bdezfouli\}@scu.edu}\\
}
}

\maketitle


\begin{abstract}
In Software-Defined Networks (SDNs), the control plane and data plane communicate for various purposes, such as applying configurations and collecting statistical data. While various methods have been proposed to reduce the overhead and enhance the scalability of SDNs, the impact of the transport layer protocol used for southbound communication has not been investigated. Existing SDNs rely on TCP (and TLS) to enforce reliability and security. In this paper, we show that the use of TCP imposes a considerable overhead on southbound communication, identify the causes of this overhead, and demonstrate how replacing TCP with QUIC can enhance the performance of this communication. We introduce the quicSDN architecture, enabling southbound communication in SDNs via the QUIC protocol. We present a reference architecture based on the standard, most widely used protocols by the SDN community and show how the controller and switch are revamped to facilitate this transition. We compare, both analytically and empirically, the performance of quicSDN versus the traditional SDN architecture and confirm the superior performance of quicSDN.

% The results confirm the lower communication overhead and message delivery delay of quicSDN versus tcpSDN.

\end{abstract}


% \fi
\IEEEpeerreviewmaketitle

\glsresetall

\section{Introduction}
\glspl{SDN} simplify new application development and facilitate network monitoring and management.
Nowadays, \gls{SDN} architectures are being used in various types of deployments, such as data center networks, \glspl{WAN} \cite{jain2013b4}, \gls{NFV} \cite{yousaf2017nfv}, 5G \cite{kaloxylos2018survey}, and edge and fog computing \cite{hu2015mobile,powell2020fog}.

The two primary components of a \gls{SDN} architecture are {controller(s)} and {switch(es)}.
A logically centralized controller implements the control plane, and switches implement the data plane.
A controller, such as Ryu \cite{ryu_controller} and \gls{ODL} \cite{odl_controller}, provides functionalities such as network topology discovery, network operation analysis, and flow rule computation and installation.
The controller provides northbound \glspl{API} to facilitate the development of various applications such as intrusion detection and load balancing.
Communication between the controller and the switches is achieved via \textit{southbound interfaces} ranging from traditional protocols such as \gls{SNMP} \cite{case1990rfc1157} to more advanced ones such as OpenFlow \cite{mckeown2008openflow}, \gls{OVSDB} \cite{pfaff2013open}, and NETCONF (Network Configuration Protocol) \cite{enns2011network}.

% In this paper, we primarily focus on OpenFlow and \gls{OVSDB}.
% These two protocols are widely deployed and are supported by off-the-shelf switches \cite{Cisco_openflow_switch, Juniper_switch} and controllers \cite{medved2014opendaylight, RYU}.
% Some vendors also provide customized versions of these protocols (e.g., Aristas' DirectFlow \cite{DirectFlow}, Cisco-OpenFlow-Plugin \cite{CiscoOpenFlowPlugin}, HP OpenFlow \cite{HPOpenFlow}).

The introduction of \gls{SDN} allows for fine-grained and centralized control over the operation of the data plane. 
For example, OpenFlow and its vendor-specific flavors such as Arista's DirectFlow \cite{DirectFlow}, Cisco-OpenFlow-Plugin \cite{CiscoOpenFlowPlugin}, HP OpenFlow \cite{HPOpenFlow}) are being used to configure flow rules and flow tables in switches via exchanging variety of messages.
The three main message types exchanged between a controller and a switch via the OpenFlow protocol are \texttt{Packet\_in}, \texttt{Flow\_mod}, and \texttt{Multipart\_request/reply}. 
When a data packet arrives at a switch and does not match the existing installed flow rules, the packet is forwarded to the controller via a \texttt{Packet\_in} message.
In large networks, such as data centers, an enormous amount of \texttt{Packet\_in} messages are generated from table misses  \cite{alsaeedi2019toward,noormohammadpour2017datacenter, curtis2011devoflow}.
This is primarily caused by the limited memory of switches and the arrival of new flows \cite{jouet2015arbitrary}.
As the table miss rate increases, the overhead of transport layer protocol used for communication between the controller and switches elevates \cite{ying2019expedited,kim2017enhanced}. 
\texttt{Flow\_mod} packets, which are used to install or modify flow rules, can be sent reactively in response to a \texttt{Packet\_in} message or proactively in anticipation of expected traffic. 
%may also be sent to switches proactively.
For example, the controller may proactively install flow rules based on the collected statistics from switches to address requirements such as load balancing.
% The size of a \texttt{Flow_mod} message depends on the flow rule complexity, and therefore, for large networks with complex rules, \texttt{Flow_mod} messages can become quite sizeable \cite{chen2021modeling}. 
%
To maintain an up-to-date view of network status, a controller needs to poll switches at regular intervals.
The controller sends \texttt{Multipart\_request} messages to each switch for each feature that it needs to collect statistics on, and each switch responds with a corresponding \texttt{Multipart\_reply} message. 
The polling frequency depends on the types of applications running on the controller.
The reply messages' size is variable and depends on the switch configuration \cite{chen2021modeling}. For instance, switches with many queues and large flow tables transmit several messages for each poll event.
Another protocol, \gls{OVSDB}, is used to configure \gls{QoS} functionalities via \gls{RPC} "transact" interactions.
Configuring a queue on a switch involves multiple \gls{OVSDB} messages: 
(i) the queue is added to the switch,
(ii) the queue is added to a specific packet scheduler, 
and (iii) the switch responds with an \gls{RPC} "update" to confirm the new configurations.
More messages are required for more complicated switch operations, and the overhead is even more significant \cite{palma2014queuepusher,caba2015apis,sharma2014implementing,flathagen2018combined,volpato2017autonomic, chen2021modeling}.

In addition to the basic functionalities offered by OpenFlow and OVSDB, over the past few years and towards the development of next-generation networks, the capability and flexibility of network switches, smartNICs, and middleboxes have considerably increased to run various configuration and management protocols.
To enhance the scalability and development of new applications and services, it is essential to provide end-to-end traffic engineering (resource allocation and security), fault detection, recovery, and isolation. 
Programmability and automation are necessary to provide such features in a scalable and dynamic fashion, and observability is essential to react to network dynamics.
To adapt to the needs of large-scale data-centers, campus networks, and carrier networks, modern switches and devices running \glspl{NOS} offer more flexibility such as programmability of data plane via P4 \cite{bosshart2014p4}, APIs for accessing and configuring switches' data plane state (e.g., OpenConfig \cite{OpenConfig2016}, NETCONF \cite{enns2011network}), event management, Linux shell access, and the capability to run containers and \glspl{VM}.
For example, Arista's \gls{EOS} \cite{eos2022} provides a \gls{SDK} for development of programs, referred to as \textit{agents}, that can access the status and configure the operation of switches. 
These agents, which can be dynamically added to or removed from the system, can implement custom protocols or rely on open protocols.
For example, \gls{gNMI} is used for configuration and telemetry, \gls{gNOI} is used for exchanging operational commands, and \gls{gRIBI} is used for exchanging commands pertaining to routing table control.
For instance, one can develop and run an agent, which continuously monitors and generates reports (including low-level counters and system temperature) that allow the controller to implement machine learning algorithms for device failure detection.

The wide range of programmability features available in today's switches reveals the increasing demand for southbound communication in \glspl{SDN}.
However, this communication imposes significant transport layer protocol overhead that severely affects the utilization of bandwidth resources and network scalability \cite{karakus2017survey,hu2014scalability}.
In this paper, we look at the communication between the control plane and data plane from a different perspective---\textit{the transport-layer protocols}.
Currently, communication reliability, packet reordering, and security in \glspl{SDN} are achieved by using \gls{TCP} (combined with \gls{TLS}).
Throughout this paper, we use the term \textit{tcpSDN} to refer to \glspl{SDN} that utilize \gls{TCP} as the transport layer protocol used for southbound communication.
tcpSDN architectures introduce various shortcomings in terms of communication overhead, lack of connection multiplexing, and high overhead of connection establishment.
For example, when multiple agents run on a switch communicating with a controller,  each agent needs to open its own connection due to the lack of supporting multiplexing connections by TCP, thereby increasing communication overhead.
Also, TCP does not deal with the \gls{HOL} blocking problem, which causes increased message delivery delay.

In this paper, we introduce \textit{quicSDN}, a novel framework to address the drawbacks and challenges of using \gls{TCP} as the transport layer protocol for southbound communication in \glspl{SDN}.
Specifically, instead of using \gls{TCP}, {quicSDN} uses a new transport layer protocol called \gls{QUIC} \cite{newQUICRFC}.
Although \gls{QUIC} was primarily designed for web traffic (HTTP3 \cite{HTTP3}), its enhancements over \gls{TCP} are applicable across various domains.
These enhancements include the ability to multiplex different streams, reduction in connection establishment latency, mitigation of the \gls{HOL} blocking problem, and adding the capability to differentiate between the ACK sent for original and retransmitted packets (a.k.a., \gls{TCP} ambiguity problem).
Towards proposing a novel architecture for \glspl{SDN}, this paper presents the following contributions:

\begin{itemize}
\item We first identify and analyze the benefits of using QUIC instead of TCP for southbound communication.
In addition to highlighting the benefits of QUIC considering the specific properties of \glspl{SDN}, we present an analytical modeling of overhead when multiple agents on a switch communicate with one or multiple agents on a controller.
In particular, this analysis shows that the overhead of QUIC drops compared to TCP as we increase the number of agents on the switch or reduce inter-message generation intervals.
Also, we justify the mitigation of \gls{HOL} problem and discuss the benefits of faster connection establishment by QUIC.
% Also, considering the short connection establishment time of QUIC, the amount of data loss on switches in the absence of connection with controllers is reduced by using QUIC.

\item Towards providing a framework for transitioning from tcpSDN to quicSDN, we discuss the design options and present a system architecture based on RYU controller and switches running \gls{OVS} and \gls{OVSDB}.
The proposed framework details aspects such as understanding and removing the intertwined dependency of RYU, \gls{OVS}, and \gls{OVSDB} on \gls{TCP} and replacing them with QUIC.
Additionally, the proposed framework details the \gls{IPC} methods to allow RYU, \gls{OVS}, and \gls{OVSDB} to communicate through QUIC.
It is worth mentioning that the proposed framework can be used to integrate additional southbound protocols (e.g., NETCONF) and controllers (e.g., OpenDaylight).
The implementation of the proposed framework is publicly and freely available.\footnote{https://github.com/SIOTLAB/quicSDN}
This implementation includes newly developed and modified entities of an \gls{SDN} architecture, which are RYU controller, \gls{OVS} switch, \gls{QUIC} protocol, and their dependent third party libraries such as Eventlet \cite{Eventlet} and Libevent \cite{provos2003libevent}.
We believe that this framework enables the research community to repeat our results and extend the functionalities of quicSDN.
Furthermore, this will lay a foundation for the evolution of \gls{QUIC} protocol in the context of \gls{SDN} architectures.

\item We built a testbed to empirically evaluate quicSDN versus tcpSDN.
We evaluate control traffic overhead versus varying message rate, loss rate, and number of streams.
The results confirm the lower communication overhead of quicSDN versus tcpSDN in all the scenarios.
Also, we measure message delivery delay between controller and switches and confirm the superior performance of quicSDN.

\end{itemize}

\vspace{5pt}

The rest of this paper is organized as follows:
Section \ref{motivation} presents an analytical study of the communication overhead of QUIC and TCP and highlights the main differences between the two protocols.
The architecture of quicSDN is presented in Section \ref{design}.
Section \ref{implementation} discusses the implementation, algorithms, and pertinent details of quicSDN.
Empirical evaluations are given in Section \ref{emp_eval}.
Section \ref{relatedWork} overviews the related work.
We conclude the paper in Section \ref{conclusion}.

\section{Motivation}
\label{motivation}
In this section, we highlight and analyze the main differences between TCP and QUIC and justify the benefits of QUIC for southbound communication in \glspl{SDN}.
For this analysis, we also develop mathematical models to showcase the benefits of quicSDN over tcpSDN.

\subsection{Communication Overhead}
\label{theor_comm_overhead}
\gls{TCP} is the widely-used transport protocol to ensure packet ordering and reliability in end-to-end message delivery.
Despite its prevalence, \gls{TCP} has several major shortcomings. 
In this section, we focus on packet transmission overhead.
\begin{figure}
\centering
\begin{subfigure}[b]{0.9\linewidth}
\includegraphics[width=\linewidth]{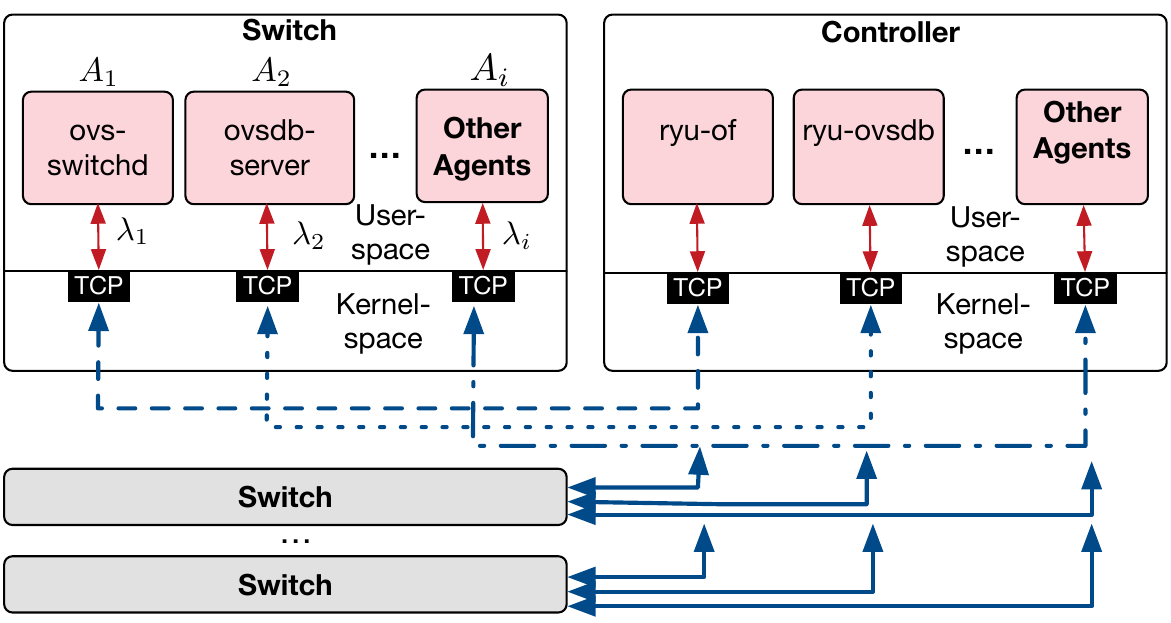}
\captionsetup{font=footnotesize}
\caption{tcpSDN Architecture.}
\label{fig:tcpSDN_arch}
\end{subfigure}
\par\bigskip
\begin{subfigure}[b]{0.9\linewidth}
\includegraphics[width=\linewidth]{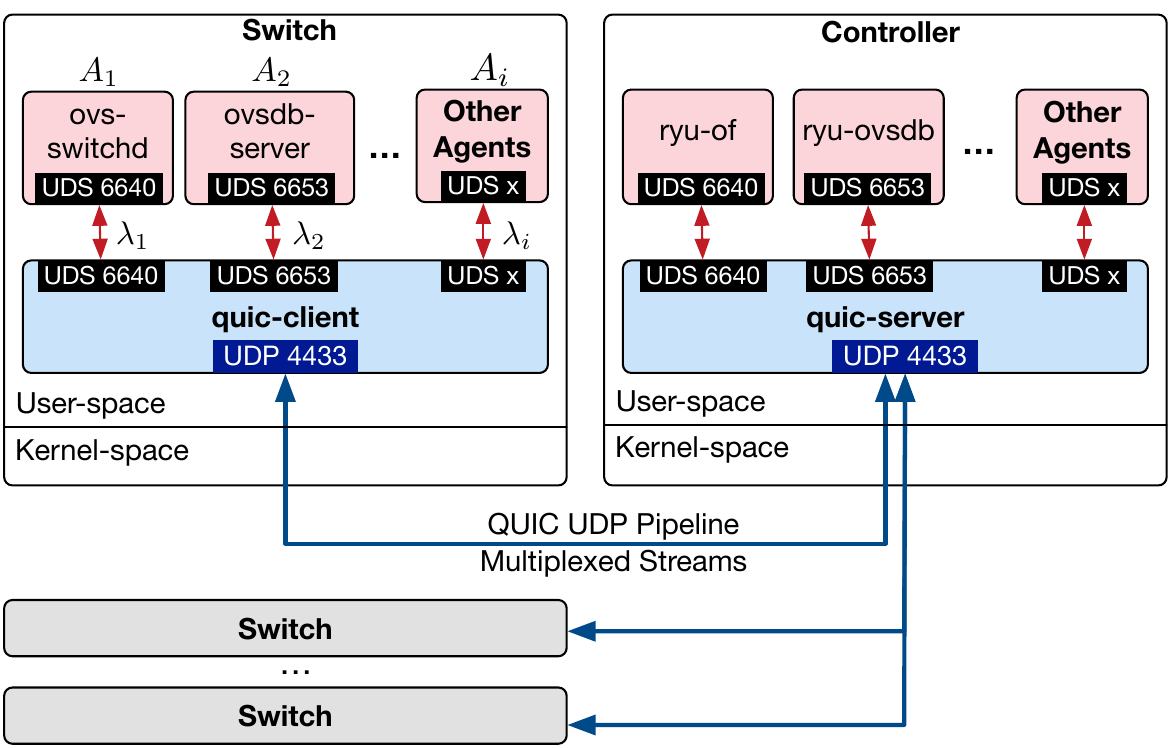}
\captionsetup{font=footnotesize}
\caption{quicSDN Architecture.}
\label{fig:quicSDN_arch}
\end{subfigure}
\captionsetup{font=footnotesize}
\caption{tcpSDN (a) and quicSDN (b) architectures. 
Compared to the tcpSDN architecture where an individual connection is required for each pair of agents (processes) running on a switch and the controller, quicSDN establishes one connection between each switch and the controller.
}
\label{fig:tcp_quic_archs}
\end{figure}

Figure \ref{fig:tcpSDN_arch} shows the tcpSDN architecture, which represents existing \glspl{SDN} that use TCP as the transport protocol for southbound communication.
Multiple \textit{agents}, a.k.a., processes or applications (such as OpenFlow and OVSDB), running on a switch need to communicate with the controller.
We denote the list of agents as $\mathbf{A} = \{A_{1}, A_{2}, ...\}$ and the number of agents as $\vert\mathbf{A}\vert$.
Using TCP, each of these agents must establish its own connection, which means the overheads pertaining to connection establishment and data exchange scale with the number of connections.
Consider the following scenario to model the overhead of data exchange between two devices.
Each agent $A_{i}$ generates message set $\mathbf{M}_{A_{i}} = \{ m^{i}_{1}, m^{i}_{2}, ...\}$, where each element of the set represents the message size in bytes.
Also, assume the intervals between message generations are short enough to place consecutive messages in a packet as long as there is available room.
Additionally, we assume the congestion window and receive window do not cause reduction in throughput, and there is no packet loss.
Using TCP, \textit{each agent must be associated with its own socket}, as Figure \ref{fig:tcpSDN_arch} shows.
% Therefore, the minimum amount of data sent from the sender to the receiver can be represented as follows:
Therefore, the minimum overhead of sending data from the sender to the receiver is:
\begin{equation}
\begin{split}
\label{eq:tcp_overhead_first}
% \mathcal{O}_{TCP} =
\sum_{\forall A_{i} \in \mathbf{A}} 
% \sum_{\forall m^{i}_{j} \in A_{i}}
((H_{EP} + H_{IP} + H_{TCP}) \times 
\\
\lceil \frac{ \sum_{\forall m_{j}^{i} \in \mathbf{M}_{A_{i}}}{m_{j}^{i}}  }{{L_{MTU}} - (H_{IP} + H_{TCP})}\rceil)
%+ 
%\sum_{j=1}^{k} {m_{j}^{i}})
\end{split}
\end{equation}
where $m_{j}^{i}$ refers to the size of message $j$ generated by agent $A_i$, $L_{MTU}$ is \gls{MTU} size, 
$H_{EP}$ is physical layer and Ethernet header size (including inter-packet gap),
$H_{IP}$ is IP header size, and $H_{TCP}$ is TCP header size (without any options field).
The values for these parameters can be found in Table \ref{header_sizes}.

\begin{table}[t]
\centering
\caption{Header size values used for the evaluations of this section.}
\label{header_sizes}
\begin{tabular}{|c|c|}
\hline
\textbf{Headers} & \textbf{Size (Byte)} \\ \hline
Ethernet and Physical Headers ($H_{EP}$)        & 28                \\ \hline
IP Header ($H_{IP}$)        & 20                \\ \hline
TCP Header ($H_{TCP}$)       & 20                \\ \hline
UDP Header($H_{UDP}$)       & 8                 \\ \hline
QUIC Short Header ($H_{QSH}$)       & 2                 \\ \hline
QUIC Frame Header ($H_{QFH}$)       & 1                 \\ \hline
\end{tabular}
\end{table}

In this paper, we propose and develop the \textit{quicSDN} architecture, demonstrated in Figure \ref{fig:quicSDN_arch}.\footnote{We will explain the implementation details in Sections \ref{design} and \ref{implementation}.}
This architecture relies on the fact that multiple agents on each switch need to communicate with a controller.
Therefore, instead of establishing individual connections between \textit{each agent and the controller}, quicSDN establishes one connection between \textit{each switch and the controller}.\footnote{The same concept applies to other data-plane components such as middle-boxes and smart NICs that need to communicate with a controller.}
The underlying QUIC protocol \textit{multiplexes} multiple connections between two endpoints and converts them into \textit{streams} inside a \gls{UDP} pipeline.
A stream is formatted as a \textit{frame} inside a packet and represents a lightweight abstraction of server-client connection and is uniquely identified by a connection ID (\texttt{cid}).
Except during the connection establishment phase, each QUIC packet includes a \gls{QSH} (denoted as $H_{QSH}$), and there is a \gls{QFH} (denoted as $H_{QFH}$) for each frame included in the packet.
%
% For the given scenario, when using QUIC, the minimum amount of data exchanged is represented as follows:
For the scenario given above, using QUIC results in the following minimum overhead:
\begin{equation}
\begin{split}
\label{eq:quic_overhead_first}
% \sum_{\forall m^{i}_{j} \in A_{i}}
(H_{EP} + H_{IP} + H_{UDP}+ H_{QSH} + \alpha \times H_{QFH})
\times 
\\
\lceil \frac{ 
\sum_{\forall A_{i} \in \mathbf{A}} 
\sum_{\forall m_{j}^{i} \in \mathbf{M}_{A_{i}}} {m_{j}^{i}}  }{{L_{MTU}} - (H_{IP} + H_{UDP} + H_{QSH} + \alpha \times H_{QFH})}\rceil 
% + 
% \\
% \sum_{j=1}^{k} {m_{j}^{i}}
\end{split}
\end{equation}
where $\alpha$ is the number of frames per packet, which depends on the size of messages and their generation pattern.

To simplify the analysis for determining the value of $\alpha$, we consider two cases, depending on the average message size. 
The maximum available space per packet for including messages is $L_{max} = L_{MTU} -( H_{EP} + H_{IP} + H_{UDP}+ H_{QSH} + H_{QFH})$.
If the average message size is larger than $L_{max}$, each packet includes either one or two frames.
For example, if the average message size is 1800 bytes, the first packet sent includes one frame (part of this message), and the second packet consists of two frames, which are the residual of the first message and the first part of the second message.
If the average message size is less than $L_{max}$, then $\alpha$ is computed as follows:
\begin{equation}
\begin{split}
\label{eq:quic_frame_per_pkt}
\mathrm{argmax}_{\alpha}
(\frac{L_{max} + H_{QFH}}{\alpha \times (m_{avg} + H_{QFH})} > 1), \;\mathrm{and}\; \alpha \in \mathbb{N}.
\end{split}
\end{equation}
%
%
% The long header has a fixed cost of 20 bytes, paid during the 1-\gls{RTT} connection establishment, and does not contribute to the overhead.
% The size of \gls{QUIC}s' short header can be small as 2 bytes; when combined with the 8 bytes of a \gls{UDP} header, \gls{QUIC} incurs 10 bytes of transport protocol overhead per packet. 

Equations \ref{eq:tcp_overhead_first} and \ref{eq:quic_overhead_first} can be used to compute overhead, neglecting inter-message generation intervals.
To represent a realistic scenario, we assume each agent ($A_{i}$) generates traffic at rate $\lambda_{A_i}$ messages per second.
This system represents $\vert \mathbf{A} \vert$ independent exponential random variables, where $\vert \mathbf{A} \vert$ is the number of agents.
To enhance the efficiency of message transmissions, transport protocols use a buffering period during which the transport layer is awaiting additional data from the application layer.
For example, Linux includes an implementation of Nagle's algorithm \cite{mogul2001rethinking}, which waits for more data from the application layer when there is pending ACK and the amount of data for transmission is less than \gls{MSS}.
% $L_{MTU} - (H_{IP} + H_{TCP})$.
We refer to the buffering delay in the transport layer as $T_{b}$.
% If not message is received during this interval, the message is sent, otherwise, the messages could be 
Therefore, to compute communication overhead, we need to model the effect of message generation burstiness arriving in the transport protocol.
Since the interval between message arrivals follows the exponential distribution, the expected number of messages generated by an agent $A_{i}$ during the buffering time is $\lambda_{A_{i}}\times T_{b}$.
%
% We compute the probability of having message bursts with varying lengths as follows.
% Considering exponential distribution, a single-message burst occurs when the inter-message interval after the current message is longer than $T_{b}$.
% Therefore, the probability of this event is $p(t > T_{b}) = e^{-\lambda}$.
% Similarly, for a two-message burst, one interval must be less than or equal to $T_{b}$ and the next interval larger than $T_{b}$.
% Therefore, $p(t > T_{b}) p(t \leq T_{b}) = e^{-\lambda}(1-e^{-\lambda})$.
% In general, the probability of an $n$-message burst is %
% \begin{equation}
% \begin{split}
% \label{eq:msg_burst_prob}
% %
% p(n) = 
% e^{-\lambda}(1-e^{-\lambda})^{n-1}
% %
% \end{split}
% \end{equation}
%
%
Assuming that all the messages are equal size ($\bar{m} = m_{1}^{i} = m_{2}^{i} = ..., \forall A_{i} \in \mathbf{A}$) and the message generation rate of all the agents is the same ($\lambda = \lambda_{A_i}, \forall A_{i} \in \mathbf{A}$), we can use message burstiness probabilities to compute the number of messages, and therefore the number of bytes generated per burst.
Let $\bar{B}_{A_i} = \{ {b}_{1}, {b}_{2}, ... \}$ represent the list of message bursts generated by agent $A_{i}$.
Each element ${b}_{j}$ is the number of bytes in a burst.
% For each agent $A_{i}$, we rely on Equation \ref{eq:msg_burst_prob} to compute the number of bytes in each burst.
% For example, denoting $\vert M_{A_{i}} \vert$ as the total size of messages generated by agent $A_{i}$,
% we compute the percentage of bytes belonging to $n$-message bursts as $p(n)\times \vert M_{A_{i}} \vert$.
The effect of burstiness on communication overhead of TCP is presented as follows:
\begin{equation}
\begin{split}
\label{eq:tcp_overhead_with_burst}
% \mathcal{O}_{TCP} =
\sum_{\forall A_{i} \in \mathbf{A}}
\sum_{\forall {b}_{j} \in \bar{B}_{A_i}}
% \sum_{\forall m^{i}_{j} \in A_{i}}
((H_{EP} + H_{IP} + H_{TCP}) \times 
\\
\lceil \frac{ {b}_{j} }{{L_{MTU}} - (H_{IP} + H_{TCP})}\rceil ).
%+ 
%\sum_{j=1}^{k} {m_{j}^{i}})
\end{split}
\end{equation}

As the message generation rate of each agent increases, the header overhead of TCP drops because data bytes belonging to different messages can be included in each packet, thereby sending larger packets.
However, QUIC performs a better job in aggregating multiple messages and sending larger packets instead of multiple smaller packets, as we show in the following.

The quicSDN architecture allows multiple agents to use a single connection to communicate with one or more processes on the controller.
Therefore, the overall rate of incoming messages into the QUIC protocol (\quicclient{ or \quicserver{} in Figure \ref{fig:quicSDN_arch}}) is higher than TCP.
Specifically, the inter-message time can be represented as the sum of $\vert \mathbf{A} \vert$ independent exponential variables. 
We represent this accumulated rate as
$ \hat{\lambda} = \sum_{\forall A_{i} \in \mathbf{A}} \lambda_{A_i}$.
%
% We will study this behavior in the performance evaluation section (\cref{result}).
With the accumulated incoming message rate $\hat{\lambda} $, we generate the list of message bursts as $\hat{\mathbf{B}} = \{ \hat{b}_{1}, \hat{b}_{2}, ... \}$, where each element $\hat{b}_{j}$ is the number of bytes in a burst.
The overhead of QUIC is computed as follows:
\begin{equation}
\begin{split}
\label{eq:quic_overhead_with_burst}
% \sum_{\forall m^{i}_{j} \in A_{i}}
\sum_{\forall \hat{b}_{j} \in \mathbf{\hat{B}}}
((H_{EP} + H_{IP} + H_{UDP} + H_{QSH} + \alpha \times H_{QFH})
\times 
\\
\lceil \frac{ 
\hat{b}_{j}
}
{{L_{MTU}} - (H_{IP} + H_{UDP} + H_{QSH} + \alpha \times H_{QFH})}\rceil).
% + 
% \\
% \sum_{j=1}^{k} {m_{j}^{i}}
\end{split}
\end{equation}

We use the models presented in this section to compare the the transmission overhead of TCP and QUIC.
Transmission overhead represents the overhead associated with the layers of the protocol stack (i.e., headers and Ethernet's inter-packet gap) when sending a certain number of messages.
% Also, we have developed a simulation tool to collect results that serve as the baselines for these evaluations.
% To ensure consistent input for analytical and simulation based comparisons, for each iteration of the experiment we use the same burst list ($\hat{\mathbf{B}}$). 
We also compute the overhead of ACK packets sent from a receiver to a sender as follows.
Neglecting the effect of packet loss and variable RTT, for a TCP connection, the number of ACK packets sent depends on the number of data packets received from the sender: the receiver either sends an ACK immediately or waits up to 500 ms to receive a second packet and then send an ACK.
Assume the number of data packets sent is denoted as $d$, the mean number of ACK packets sent is $(d+d/2)/2$.
Each ACK packet includes a TCP header, in addition to the headers of underlying layers.
A similar method is used to compute the ACK overhead of QUIC.

\begin{figure*}
     \centering
     \begin{subfigure}[b]{0.24\textwidth}
         \centering
         \includegraphics[width=\textwidth]{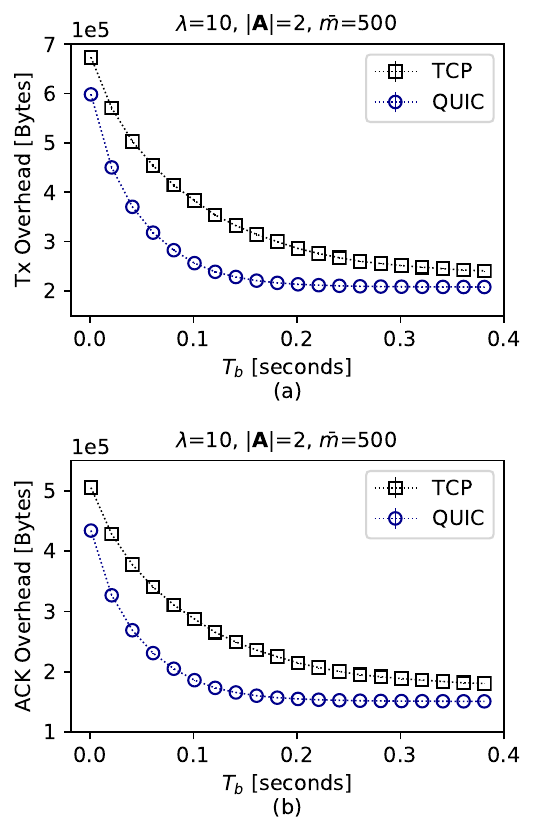}
        %  \caption{$y=x$}
        %  \label{fig:y equals x}
     \end{subfigure}%
     \begin{subfigure}[b]{0.24\textwidth}
         \centering
         \includegraphics[width=\textwidth]{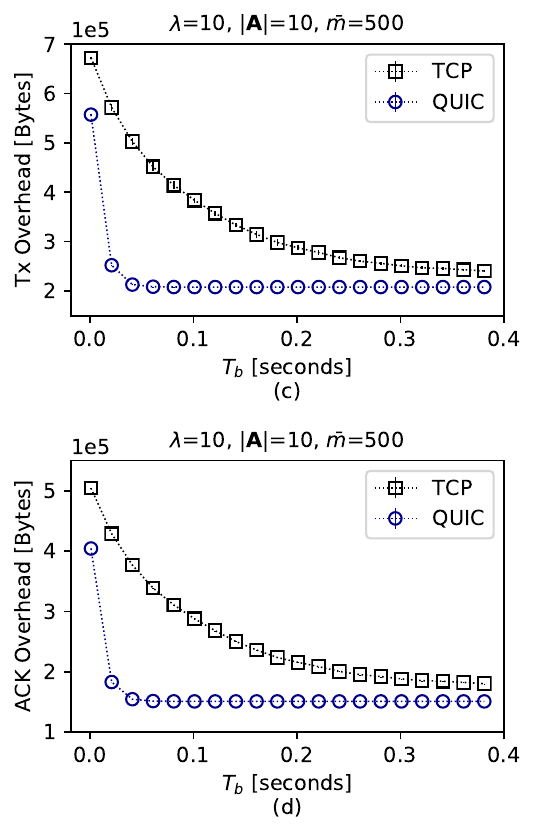}
        %  \caption{$y=3sinx$}
        %  \label{fig:three sin x}
     \end{subfigure}%
     \begin{subfigure}[b]{0.24\textwidth}
         \centering
         \includegraphics[width=\textwidth]{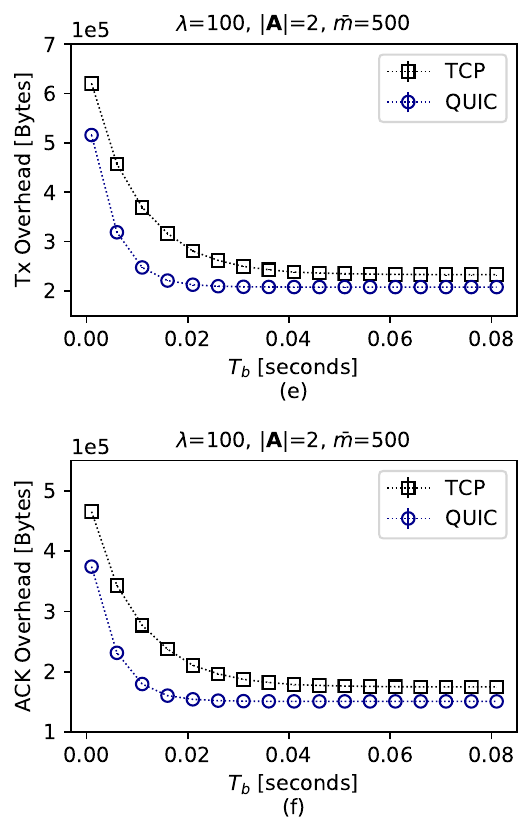}
        %  \caption{$y=5/x$}
        %  \label{fig:five over x}
     \end{subfigure}%
     \begin{subfigure}[b]{0.24\textwidth}
         \centering
         \includegraphics[width=\textwidth]{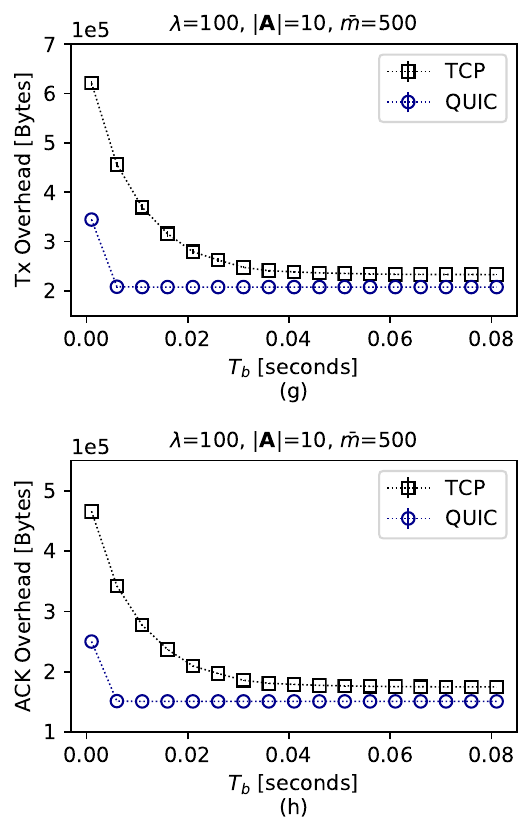}
        %  \caption{$y=5/x$}
        %  \label{fig:five over x}
     \end{subfigure}
    \captionsetup{font=footnotesize}
    \caption{Each column shows the transmission overhead and acknowledgement overhead of TCP and QUIC in various scenarios. 
    The first row shows transmission overhead and the second row shows acknowledgment overhead.
    $\lambda$ is rate of message generation by each agent.
    $\vert\mathbf{A}\vert$ is the number of agents. $\bar{m}$ is average message size.
    These results confirm the considerably lower packet exchange overhead of QUIC versus TCP.
    }
    \label{fig:500b_tcp_vs_quic}
\end{figure*}

Figure \ref{fig:500b_tcp_vs_quic} presents the results when we vary message rate ($\lambda$) and the number of agents ($\vert \mathbf{A} \vert$).  
The average message size ($\bar{m})$ is 500 bytes. 
The total number of messages generated per agent is 5000.
These results show that as the buffering delay ($T_{b}$) increases, the difference between the overhead (and the number of packets) of TCP and QUIC reduces until they reach their minimum values.
The lower acknowledgment overhead of QUIC is a direct effect of a lesser number of packets sent by this protocol.
Since QUIC can multiplex multiple agents' messages into one connection, its communication overhead declines and stabilizes faster than TCP.
Especially, as the number of agents increases, the decline rate of QUIC's overhead increases as well, and this can be observed by comparing the first column with the second column and the third column with the fourth column.
%
% For example, while the overhead of QUIC in sub-figure (a) is up to x\% lower than TCP, the reduction increases to y\% in sub-figure (c).

It must be noted that the multiplexing feature of quicSDN comes at a cost, associated with attaching \gls{QFH} to each frame in a packet.
Specifically, the overhead of QUIC is higher than TCP when $H_{UDP} + H_{QSH} + \alpha \times H_{QFH} > H_{TCP}$.
%
% \begin{equation}
% \begin{split}
% \label{eq:tcp_overhead}
% %
% H_{UDP} + H_{QSH} + \alpha \times H_{QFH} > H_{TCP}
% %
% \end{split}
% \end{equation}
%
For example, considering the parameters given in Table \ref{header_sizes}, the overhead of quicSDN is higher than tcpSDN when more than ten streams are included in a packet.
Nevertheless, quicSDN results in a lower number of packet transmissions by establishing a single connection to carry the data of all the agents.
We will empirically evaluate the overhead of quicSDN and tcpSDN in Section \ref{empirical_overhead_eval}.

\subsection{Head of Line Blocking Problem}

\begin{figure}[t]
    \centering
    \includegraphics[width=1.0\linewidth]{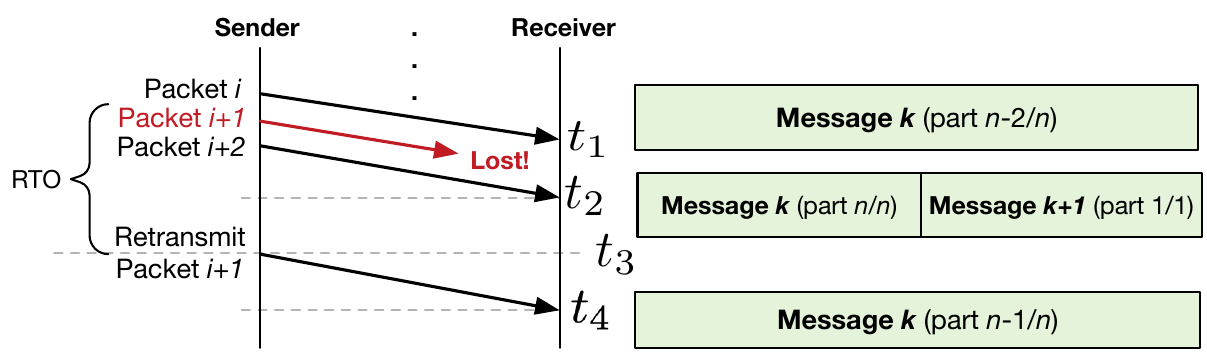}
    \captionsetup{font=footnotesize}
    \caption{The effect of \gls{HOL} blocking on message delivery delay to application layer.
    In this scenario, the delivery of Message $k+1$ is at least $RTT + RTT/2$ slower when using TCP, compared to QUIC.
    }
    \label{fig:HOL}
\end{figure}

The use of multiplexing allows QUIC to mitigate the \gls{HOL} blocking problem.
In TCP, when packets arrive out of order at the receiver, since the protocol is unaware of the boundary between messages, it cannot deliver completely received messages to the application layer.
Whereas, QUIC reduces message delivery delay by assigning messages to streams.
% we could include a sample scenario and show how increasing RTT results in longer delay with TCP
%
For example, consider the scenario given in Figure \ref{fig:HOL}, where a controller (sender) sends two messages, Message $k$ and Message $k+1$ to a switch (receiver).
The transmission of Message $k$ is performed by sending $n$ packets.
Packet $i$ includes part $n-2$ of Message $k$ and is successfully received by the receiver at time $t_1$.
Packet $i+1$, which includes part $n-1$ of Message $k$, is lost.
Packet $i+2$ includes the last part of Message $k$ and all the bytes of Message $k+1$.
At time $t_{2}$, although Message $k+1$ has been fully received, TCP does not deliver it to the application layer.
In contrast, QUIC can use frame headers to identify message boundaries, and therefore, Message $k+1$ is delivered to the application layer as soon as Packet $i+2$ is received.
The retransmission of Packet $i+1$ is triggered by the \gls{RTO} of sender; alternatively, assuming that more packets are sent after packet $i+2$, the reception of three duplicate ACKs triggers the TCP fast retransmission method to retransmit Packet $i+1$.
In either case, the delivery delay of Message $k+1$ to the application is at least one \gls{RTT} delayed when using TCP, compared to QUIC.
The shorter message delivery delay of QUIC is beneficial in \glspl{SDN}.
For example, if Message $k+1$ is a flow rule, quicSDN provides faster reaction to new flow arrival into a switch.
We will empirically evaluate the effect of \gls{HOL} in Section \ref{message_delivery_delay}.

% Specifically, this problem occurs when the segments before a missing segment can be processed to form messages \cite{scharf2006nxg03, qian2015tm3}. %

\subsection{Connection Establishment and Migration}
Both TCP and QUIC need to establish a connection before data exchange.
% \subsubsection{Connection Establishment and Migration}
% In tcpSDN, when a switch is assigned to a new controller for load balancing purposes, the controller and switch need to establish new \gls{TCP} connections.
% When the switch-to-controller connections are short-lived, there will be a significant connection establishment overhead.
%
\gls{TCP} with \gls{TLS}1.2 \cite{dierks2008transport} and \gls{TLS}1.3 \cite{TLS1_3} require three and two \gls{RTT}s, respectively, for connection establishment.
By leveraging a multi-stage key exchange, \gls{QUIC} combines the transport and security layer connection establishment procedures to minimize connection establishment overhead to one \gls{RTT}.
In the first stage, the client sends a 'hello' message (\texttt{CHLO}) to retrieve the server's configuration.
Since the client is unknown to the server, the server responds with a \texttt{REJ} packet.
The \texttt{REJ} packet contains the server's configuration, long term  Diffie-Hellman value, key agreement, \texttt{cid}, and initial data.
The client then authenticates the server by verifying the certificate chain and signature.
After authentication, the client sends a complete \texttt{CHLO} packet to the server and finishes the first handshake.
At this stage, the client has the initial keys and is ready to exchange application data with the server.
% In one \gls{RTT} connection establishment, the client sends application data to the server by using the initial keys, even before receiving a reply from the server.
Upon a successful first handshake, the server sends a complete hello (\texttt{SHLO}) to the client and concludes the final handshake.
% Apart from the initial handshake packets, \gls{QUIC} packets are fully authenticated and partially encrypted.
% The non-encrypted part of the packet is used for routing and is also used to decrypt the remaining part of the packet.
To support connection migration, \gls{QUIC} uses a unique \texttt{cid} to identify each connection.
This allows for connection rebinding even if the connection parameters such as IP address or port number are changed.
% Typically, the server requests a \texttt{cid} for the lifetime of the connection.
Therefore, if a switch is assigned to a controller that it has communicated with  in the past, quicSDN can establish the connections in  zero \gls{RTT}, while \gls{TCP} with \gls{TLS}1.3 requires one \gls{RTT}.

\begin{figure}[t]
    \centering
    \includegraphics[width=1.0\linewidth]{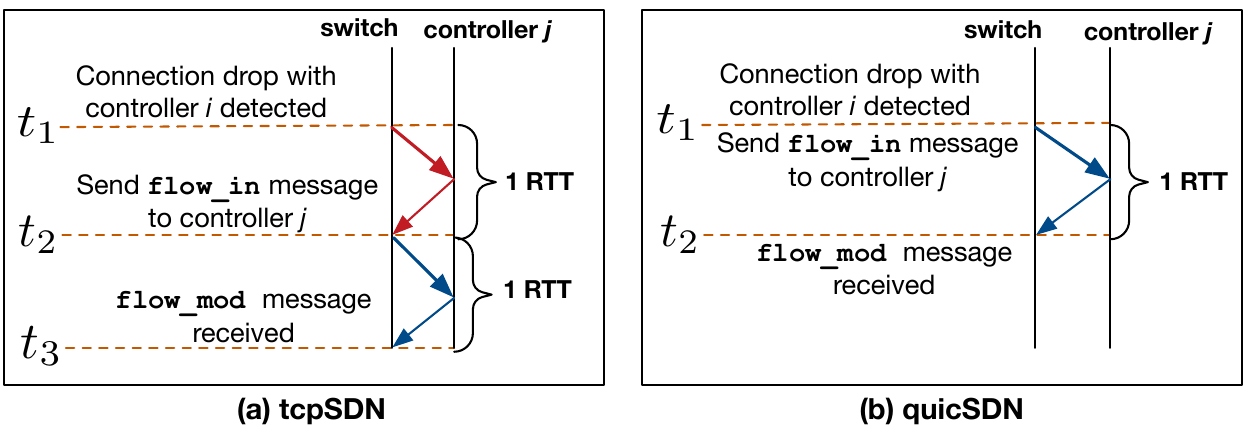}
    \captionsetup{font=footnotesize}
    \caption{Connection reestablishment delay between switches and controllers in (a) tcpSDN and (b) quicSDN.
    quicSDN facilities dynamic assignment of switches to controllers with a shorter delay and lower overhead.
    }
    \label{fig:conn_reestablishment}
\end{figure}

The shorter connection establishment time of QUIC allows quicSDN to provide the following benefits.
First, if a switch detects connection drop with a controller, the delay of connection reestablishment with the same or another controller is lower compared to tcpSDN.
A sample scenario is given in Figure \ref{fig:conn_reestablishment}.
Assume at time $t_{1}$ the switch detects connection drop with controller $i$.
This is detected, for example, when the switch sends a \texttt{flow\_in} to the controller, but no response is received within a timeout period.
At this point, if using tcpSDN, the switch needs to first establish a connection with controller $j$ (during $t_1$ to $t_2$).
Then, at time $t_2$, the \texttt{flow\_in} message is sent to controller $j$.
This process requires 2$\times$RTT.
In contrast, quicSDN eliminates the connection reestablishment delay and immediately sends the \texttt{flow\_in} message to controller $j$ at time $t_1$. Thereby, the delay of communication with controller $j$ is reduced to RTT.
The lower delay of connection reestablishment in quicSDN also reduces the overhead of (proactive) switch reassignment to controllers (e.g., when the load of controllers are periodically balanced).
Therefore, the quicSDN framework proposed in this paper facilitates the development of enhanced load balancing and controller assignment solutions, building on top of the methods proposed in existing works \cite{bera2020dynamic,muller2014survivor,alowa2019combined,qin2018SDNController}.

\subsection{Congestion and Flow Control}
\gls{QUIC}'s congestion control mechanism provides a richer set of features compared to \gls{TCP} \cite{newQUIC10000}. 
For instance, consider the \gls{TCP} ambiguity problem, where \gls{TCP} cannot determine if the ACK was for the original or retransmitted packet.
\gls{QUIC} solves this problem by assigning a unique \textit{Packet Number} to each packet, irrespective of being an original or a retransmission.
%This unique sequence number is called \textit{Packet Number}.
\gls{QUIC} also reduces congestion control by using a \gls{NACK} scheme, where, instead of acknowledging every packet, the receiver notifies the sender about lost packets \cite{QUICLoss}.

In \gls{TCP}, a sender can be blocked from sending data when the entire receiver buffer is consumed.
\gls{QUIC} addresses this problem via two methods:
First, with connection-level flow control, in which an upper-limit is imposed on the entire connection for a receiver's aggregated buffer.
Second, flow-level flow control imposes an upper-limit on the flow-level buffer size on the receiver.
To reduce or increase flow-level buffer size, \gls{QUIC} uses a window update frame for advertising per-stream absolute byte offsets for received, delivered, and sent packets.
These per-stream absolute byte offsets dictates the amount of bytes a receiver is willing to accept on a particular stream.

\section{quicSDN Architecture}
\label{design}
This section presents a high-level overview of the interactions between the components of quicSDN:
\gls{QUIC}, \gls{OVS}, and RYU. 
In typical scenario, \gls{QUIC} is used by an application by incorporating \gls{QUIC}'s code into the application's code and compiled as one agent.
This prohibits the use of single \gls{QUIC} instance by multiple applications.
The memory allocation performed for queues and buffers of \gls{QUIC} is restricted to the application it was compiled with.
The quicSDN architecture is different than this method because, on a device (switch or controller), multiple agents (processes) interact with a \gls{QUIC} instance.
As Figure \ref{fig:quicSDN_arch} shows, \ovsdbserver{}, \ovsswitchd{}, and \quicclient{} run on the switch, and \ryuovsdb{}, \ryuopenflow, and \quicserver{} run on the controller. 
\ovsdbserver{} and \ovsswitchd{} are agents responsible for processing \gls{OVSDB} and OpenFlow packets on the switch side. 
\ryuovsdb{} and \ryuopenflow{} are agents running on the controller to process the \gls{OVSDB} and OpenFlow packets respectively.
\quicclient{} and \quicserver{} are agents running on the switch and controller, respectively, to establish communication between the switch and the controller.
%
%
% \begin{figure}[t]
%     \centering
%     \includegraphics[width=1.0\linewidth]{overall_architecture.pdf}
%     \captionsetup{font=footnotesize}
%     \caption{The overall software architecture implemented by the switches and controllers in a quicSDN architecture. 
%     }
%     \label{fig:OverallArchiSoft}
% \end{figure}

\subsection{Inter-process Communication (IPC)}
Since \gls{QUIC} is an application-layer protocol, it cannot be used as an operating system's inbuilt transport protocol (like \gls{TCP} or \gls{UDP}).
Therefore, an \gls{IPC} is required to facilitate communication between \gls{QUIC} and application processes.
This section describes the pros and cons of various \gls{IPC} methods for quicSDN.

\subsubsection{Shared Memory}
To allow multiple applications (agents) to communicate through shared data structures, either they must be complied as a single application, or they can use shared memory via a memory map.
One of the main drawbacks of these methods is the lack of extensibility and abstraction.
Specifically, accessing the source code of all the modules is necessary to implement these methods. 
For example, suppose there is a plan to extend a switch's features by adding a component; in that case, its code must be fully available to be integrated with the existing ones.
Also, even when the new component's source code is available, the developer still needs to understand the execution paths of the code thoroughly.
For instance, code modification and the introduction of new threads are usually required to allow concurrent execution of components.
Furthermore, the larger code size and the lack of clear interfaces between modules result in more complicated code debugging and enhancements when employing these methods.
Additionally, accessing shared data structures also causes race conditions.
It is essential to acquire mutually exclusive locks to avoid race conditions among message producers and consumers.
These locks can cause performance bottleneck by introducing differences in the rate of packets processed by switches or controller.
This observation has been made in multiple studies \cite{odaira2003selective,jung2014scalable}.

\subsubsection{Message Passing}
Compared to shared memory, message passing methods are easier to implement and more extensible. 
The two primary methods of message passing are message queues and \gls{UDS}.
To simplify system extensibility, we use the latter method because of its ease of debugging and support in all the major programming languages.
There are two primary types of \glspl{UDS} at the transport layer level: stream sockets and datagram sockets.
With stream \gls{UDS}, received data is in form of stream bytes. 
The arrival of stream bytes can be out-of-order and it is required to be put in order by the application based on message boundaries.
Finding message boundaries in stream sockets introduces processing overhead because they are byte oriented and the receiver needs to parse and rearrange the received bytes, thereby introducing additional overhead.
On the other hand, datagram sockets are faster and allow an entire message to be passed.
% This obviates the need for message boundary detection and can be used for implementing various message transmission scheduling methods.

\subsection{Switch}
\label{design_switch}
\begin{figure}[t]
    \centering
    \includegraphics[width=1\linewidth]{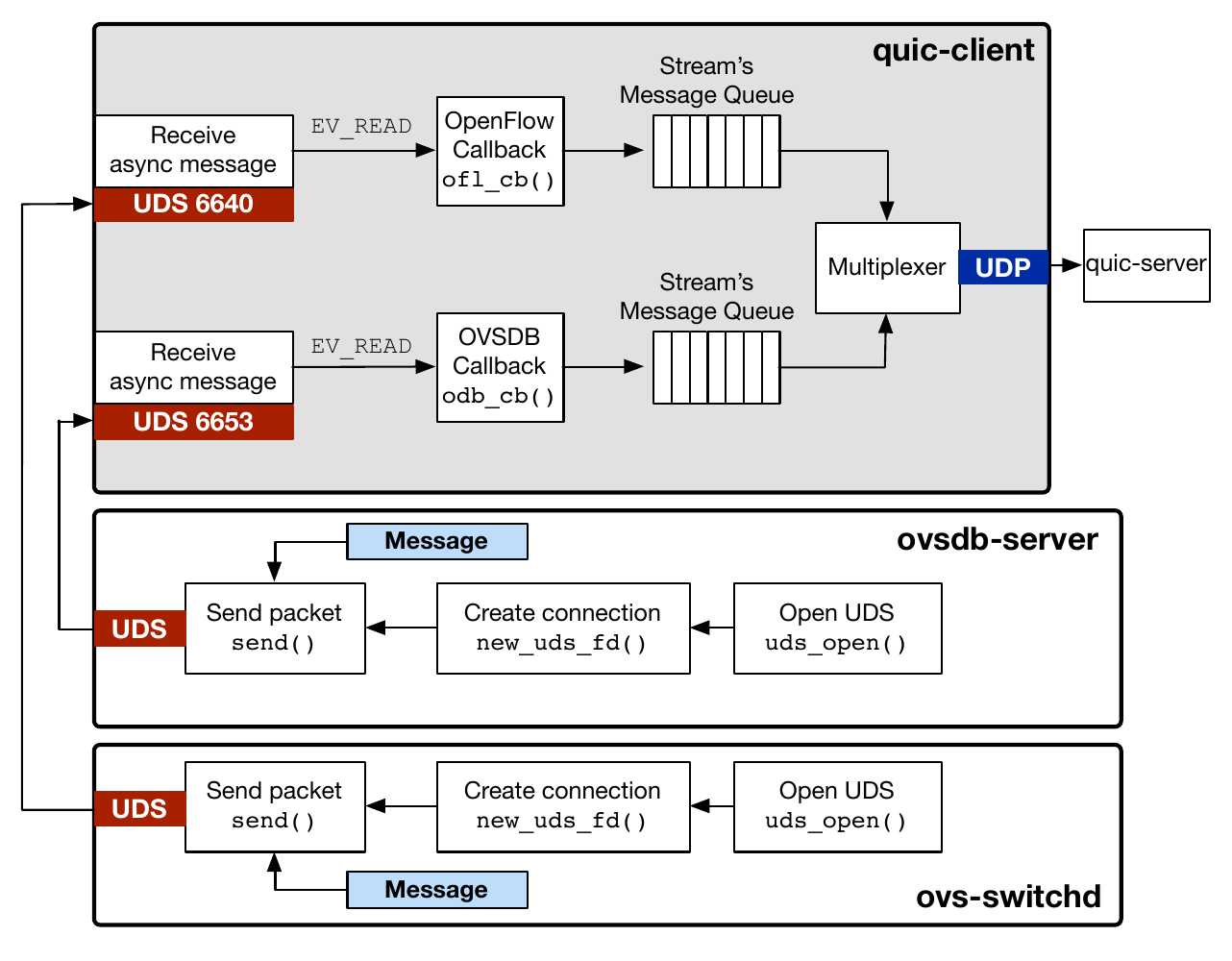}
    \captionsetup{font=footnotesize}
    \caption{The architecture of a switch in quicSDN. 
    The figure highlights the modifications to \gls{OVS} and how packets are processed by the \quicclient{} component. 
    }
    \label{fig:OVSSideQUIC}
\end{figure}

Within the quicSDN architecture, multiple processes on a switch can communicate with the controller via the \quicclient{} module.
There are two entities that communicate with \quicclient{}: 
\ovsswitchd{} and \ovsdbserver{}, handling OpenFlow and OVSDB, respectively.
Figure \ref{fig:OVSSideQUIC} presents the architecture of a quicSDN switch. 

On tcpSDN switches, \ovsswitchd{} and \ovsdbserver{} are connected to the controller using the following \gls{CLI} commands:
\begin{itemize} \footnotesize
    \setlength{\itemindent}{-1em}
    \item  [\texttt{\$}] \texttt{ovs-vsctl set-controller <bridge name> tcp:<controller-IP>:<port>}
    \item  [\texttt{\$}] \texttt{ovs-vsctl set-manager tcp:<controller-IP>:<port>}
\end{itemize}
\noindent
The quicSDN architecture provides new \gls{CLI} commands to allow \ovsdbserver{}'s and \ovsswitchd{}'s \glspl{UDS} to communicate with \quicclient{}.
\begin{itemize} \footnotesize
    \label{cli_udp}
    \setlength{\itemindent}{-1em}
    \item  [\texttt{\$}] \texttt{ovs-vsctl set-controller <bridge name> udp:<controller-IP>:<port>}
    \item  [\texttt{\$}] \texttt{ovs-vsctl set-manager udp:<controller-IP>:<port>}
\end{itemize}
To enable the new interface, we developed the \texttt{udp\_vconn\_class} class and its associated function pointers to search for the "udp" keyword in the \glspl{CLI} and open the \gls{UDP} connections to \quicclient{}. 
The opened connections are mapped to stream pointer \glspl{FD}, which are registered in function \texttt{new\_uds\_fd()}.
The aforementioned process is for both \ovsdbserver{} and \ovsswitchd{}.

\quicclient{} spawns two \gls{UDP} servers listening on ports 6653 and 6640. 
The messages received on these ports are processed and multiplexed in \quicclient{} and then transmitted to the \quicserver{}. 
To avoid any thread blockage while waiting for packet arrivals, we use async I/O operations by leveraging the \textit{libevent} library, a concurrent, highly scalable network library.
Libevent library attaches a callback function to a \gls{FD} associated to an application.
This callback function is invoked and notifies the application if an event occurs on the \gls{FD}, such as receiving or sending data.
The two newly-introduced \glspl{FD} for sockets, along with their callbacks, on ports 6653 and 6640 are mapped to stream pointers in \quicclient{} to communicate with \ovsdbserver{} and \ovsswitchd{}. \label{streamID}
%The two callbacks associated with these \glspl{FD} are used to detect activity on the sockets. 
The \gls{QUIC} RFC \cite{newQUICRFC} mandates the use of even and odd stream IDs for client-initiated and server-initiated connections, respectively.
In order to distinguish packets received on ports 6653 and 6640 on \quicclient{}, different stream IDs are selected for OpenFlow and OVSDB.
Since all stream IDs of client-initiated connections in \quicclient{} are even, we assign even stream IDs divisible by 3 for messages sent to \openflowryuapp{} and the rest are used for messages sent to \ovsdbryuapp{}.
This assignment happens in the round robin fashion for the available opened streams.
Packets generated on the stream IDs are multiplexed into the same \gls{UDP} pipeline for transmission to \quicserver{}.

\subsection{Controller}
\begin{figure}[t]
    \centering
    \includegraphics[width=1 \linewidth]{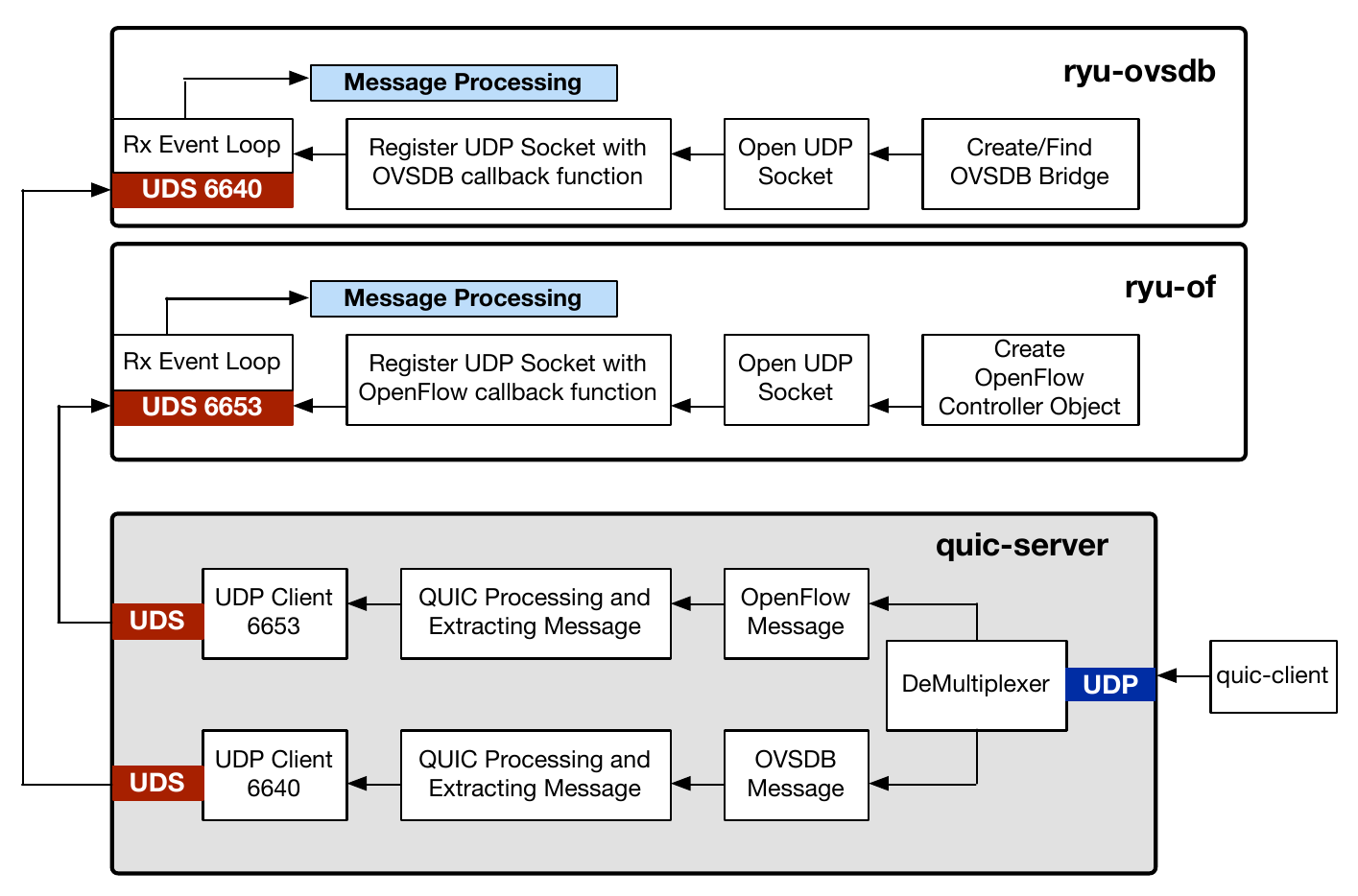}
    \captionsetup{font=footnotesize}
    \caption{The architecture of  controller in quicSDN. This figure highlights the modifications to RYU and shows how packets are processed by the \quicserver{} component.
    }
    \label{fig:RYUSideQUIC}
\end{figure}

Figure \ref{fig:RYUSideQUIC} shows the controller architecture.
The two main entities of quicSDN's controller are \quicserver{} and RYU.
The RYU entity includes two agents, \openflowryuapp{} and \ovsdbryuapp{}, which communicate with \quicserver{} over a datagram connection on ports 6653 and 6640.

The RYU controller's asynchronous I/O infrastructure is based on the \textit{eventlet} library, which is a highly scalable and non-blocking I/O library.
The eventlet library socket implementation is different than the standard \texttt{socket.socket} class in Python.
The eventlet library implements sockets as GreenSockets \cite{GreenSocket} and sets them into a non-blocking state to support asynchronous I/O operations.
RYU spawns a server based on GreenSocket and registers an event loop to receive data on the GreenSocket.
In order to make it \gls{UDP} compatible, the event loop is modified by dismantling all the \gls{TCP} related code and modifying the callbacks.

After receiving packets from \quicclient{}, \quicserver{} performs demultiplexing by disassembling streams based on their IDs.
If the stream ID is divisible by 3, then the packet is deliverd to \openflowryuapp{}, otherwise it is delivered to \ovsdbryuapp{}.

\section{Implementation}
\label{implementation}
This section presents the newly developed and modified entities in quicSDN. 
% Furthermore, this section details the steps needed the transition from a tcpSDN to quicSDN architecture.
%
We use color-coding schemes to highlight the newly-developed and modified entities.
Blue highlights indicate newly-developed entities, and the red highlights indicate modified entities.
We use three programming languages in our implementations: 
C++ for \gls{QUIC}, Python for Ryu, and C for \gls{OVS}.
We use these languages to meet the varying performance and programmability requirements of different \gls{SDN} software components. 
For example, we use Python to program the controller because it simplifies application development and extensibility.
On the switch, we use C language because our main focus is to enhance performance.
   
\subsection{OVS}   
This section describes the modifications made to \gls{OVS} to make it compatible with quicSDN.
With tcpSDN, the OpenFlow and \gls{OVSDB} protocols used by \gls{OVS} are implemented on 
\gls{TCP}, sitting in the Linux kernel network stack.
The transport layer parameters are defined in the \texttt{rconn} C structure (struct) inside the \gls{OVS} code.
There is one \texttt{rconn} C structure per transport connection between the controller and the switch.
For instance, the OpenFlow and \gls{OVSDB} connections use different \texttt{rconn} C structures, even though the endpoints are the same.
The \texttt{rconn} C structure is used to maintain socket, port, and other protocol-related information. 
To support quicSDN, the \texttt{rconn} C structure needs to be modified to support \gls{UDP}.
The OpenFlow and \gls{OVSDB} protocols are implemented as service objects inside the \gls{OVS} code.
Each service object is an abstract protocol process.
For instance, in order to start the OpenFlow and \gls{OVSDB} services, \gls{OVS} creates a service object for each service, and each service object is tied to its \texttt{rconn} C structure.

The OpenFlow and \gls{OVSDB} services in \gls{OVS} are created by issuing the \gls{CLI} commands presented in \cref{cli_udp}. %and specifying the service parameters.
While an \gls{OVS} service is created, the \texttt{vconn\_lookup\_class()} function looks up the requested transport protocol in the \gls{CLI} against a list of predefined connection classes. 
The \gls{UDP} connection class \texttt{udp\_vconn\_class} inherits the connection-related function pointers, including \texttt{open()}, \texttt{close()}, \texttt{connect()}, \texttt{recv()}, and \texttt{send()}.
The \texttt{open()} function establishes a connection with the controller and must not be blocked while waiting for the connection requests or responses.
If connection establishment does not complete immediately, the socket returns \texttt{EINPROGRESS} and retries in the background. 
\texttt{close()} tears down the connection gracefully, \texttt{send()} sends, and \texttt{recv()} receives OpenFlow messages.
Similar to \texttt{open()}, \texttt{recv()} does not block while waiting for the messages to arrive.

Using the newly modified transport layer infrastructure, the function \texttt{new\_udp\_uds()} opens a \gls{UDP} socket for each \gls{OVS} service.
These sockets are registered to the new \glspl{FD} in function \texttt{new\_uds\_fd()}, which are attached to their function pointers \texttt{open()}, \texttt{close()}, \texttt{recv()}, and \texttt{send()}.
At this point, the \texttt{rconn} C structures are populated and the \gls{OVS} services enter into the \connecting{} state.
The OpenFlow and \gls{OVSDB} services' state are dependent on the underlying transport layer protocol. 
Since there are no state transitions in \gls{UDP} to show whether the connection is in an established state or not, the \gls{OVS} services immediately transitions into the \activeState{} state.
\begin{algorithm}[!htb]
    
	\footnotesize
	\SetInd{0.9em}{0.7em}
	 \caption{Pseudo-code of {\quicclient{}}}
	\label{alg:QUIC_CL_CODE}
    \SetKwFunction{FMain}{main}
    \SetKwProg{Fn}{function}{}{}
    \Fn{\FMain{}}
    {
        \textcolor{blue_color}{p\_openflow, p\_ovsdb}, p\_quic = \{sock, port, addr\} \\
        client\_arg = \{p\_openflow, p\_ovsdb, p\_quic\} \\
        \textcolor{red_color}{populate client\_arg from CLI} \\
        \textcolor{blue_color}{p\_openflow = Connect to \ovsswitchd{} on port 6653} \\
        \textcolor{blue_color}{p\_ovsdb = Connect to \ovsdbserver{} on port 6640} \\
        
        \uIf {!($start\_client$(client\_arg)} { 
            return $failure$ \\
        } 
            
        \textcolor{blue_color}{close(p\_openflow, p\_ovsdb, p\_quic)}\\
        \KwRet\
    }
        
        \SetKwFunction{FMain}{start\_client(client\_arg)}
        \SetKwProg{Fn}{function}{}{}
        \Fn{\FMain{}}
        {
            %int sock = -1; \\
            \textcolor{blue_color}{s1 = client\_arg$\rightarrow$p\_openflow$\rightarrow$sock} \\
            \textcolor{blue_color}{File *fp\_ofl
            = \texttt{fileno(s1)}} \\
            \textcolor{blue_color}{s2 = client\_arg$\rightarrow$p\_ovsdb$\rightarrow$sock} \\
            \textcolor{blue_color}{File *fp\_odb 
            = \texttt{fileno(s2)}} \\

            sock = create UDP socket to connect to \quicserver{} \\
            File *fd\_ = \texttt{fileno(sock)} \\
            \texttt{// set event callbacks} \\
            fd\_ $\rightarrow$ \texttt{readcd(), writecb()} \\
            \textcolor{blue_color}{fp\_ofl $\rightarrow$ \texttt{ofl\_cb()}} \\
            \textcolor{blue_color}{fp\_odb $\rightarrow$ \texttt{odb\_cb()}} \\
            
            \_quic = \texttt{init()} \\
            \uIf {!($\_quic\rightarrow run(client\_arg)$} { 
                return $failure$ \\
            }
            \KwRet\
        }
        
         \SetKwFunction{FMain}{init}
        \SetKwProg{Fn}{function}{}{}
        \Fn{\FMain{}}
        {
            \_quic$\rightarrow$client = Initialize new client \\
            set \textcolor{red_color}{fd\_}, \textcolor{blue_color}{fp\_ofl}, \textcolor{blue_color}{fp\_odb} event callbacks in \_quic\\
            \KwRet \_quic\
        }

        \SetKwFunction{FMain}{run}
        \SetKwProg{Fn}{function}{}{}
        \Fn{\FMain{client\_arg}}
        {
            \uIf{(session\_file)}{
            \tcp{0-RTT Scenario}
                \uIf{!(resume())}{
                    \KwRet $failure$\
                }
            }
            \Else{
                \tcp{1-RTT Scenario}
                \texttt{do\_handshake()} \\
                \uIf{!(connect()}{
                    \KwRet $failure$\
                }
            }
            
            \tcp{Starting event loop }
            \texttt{ev\_run(ev\_d, 0)} \\
            \KwRet\
        }
         \SetKwFunction{FMain}{readcb}
        \SetKwProg{Fn}{function}{}{}
        \Fn{\FMain{ev\_loop *loop, ev\_io *w}}
        {
            auto c = <client *>w$\rightarrow$data; \\
            \textcolor{red_color}{\texttt{on\_read()}} \\
        }  
        
        \SetKwFunction{FMain}{writecb}
        \SetKwProg{Fn}{function}{}{}
        \Fn{\FMain{ev\_loop *loop, ev\_io *w}}
        {
            auto c = <client *>w$\rightarrow$data \\
            \textcolor{red_color}{\texttt{on\_write()}} \\
        } 
        \SetKwFunction{FMain}{\textcolor{blue_color}{ofl\_cb}}
        \SetKwProg{Fn}{function}{}{}
        \Fn{\FMain{ev\_loop *loop, ev\_io *w}}
        {
            \textcolor{blue_color}{auto c = <ofl *>w$\rightarrow$data}\\
            \textcolor{blue_color}{this$\rightarrow$type\_flag = openflow} \\
            \textcolor{blue_color}{\texttt{on\_ofl\_odb\_read()}} \\
        } 
        
        \SetKwFunction{FMain}{\textcolor{blue_color}{odb\_cb}}
        \SetKwProg{Fn}{function}{}{}
        \Fn{\FMain{ev\_loop *loop, ev\_io *w}}
        {
            \textcolor{blue_color}{auto c = <ofdb *>w$\rightarrow$data} \\
            \textcolor{blue_color}{this$\rightarrow$type\_flag = ovsdb} \\
            \textcolor{blue_color}{on\_ofl\_odb\_read();} \\
        } 
    
\end{algorithm}

\begin{algorithm}[!htb]
    
	\footnotesize
	\SetInd{0.9em}{0.7em}
	    \caption{Pseudo-code of {\quicclient{}} (continued from Algorithm 	\ref{alg:QUIC_CL_CODE})}
	\label{alg:QUIC_Client_callbacks}

        \SetKwFunction{FMain}{on\_read}
        \SetKwProg{Fn}{function}{}{}
        \Fn{\FMain{}}
        {
            array<uint8\_t, 65536> buf \\
            \While{true}{
                \uIf{!(recvfrom(this$\rightarrow$fd\_, buf.data, buf.len))}{
                        \KwRet $failure$\
                }
                \uIf{!(feed\_data(buf.data, buf.len)}{
                    \KwRet $failure$\ 
                }
            }
        }  
        \SetKwFunction{FMain}{feed\_data}
        \SetKwProg{Fn}{function}{}{}
        \Fn{\FMain{uint8\_t data, int data\_len}}
        {
            \uIf{handshake\_completed}{
                 \uIf{!\texttt{\textcolor{red_color}{\_con\_recv(data, datalen, \&stream\_id)}}}{
                        \KwRet $failure$\   
                 }
                 \uIf{stream\_id is divisible by 3}{
                    \texttt{sendto($this\rightarrow fp\_ofl$)}
                 }
                 \Else{
                    \texttt{sendto($this\rightarrow fp\_odb$)}
                 }
            }
            \Else{
                \uIf{!$do\_handshake(data, datalen$)} {
                    \KwRet $failure$\
                }
                \Else{
                    handshake\_completed = true
                }
            }
            \KwRet\
        }  
        \SetKwFunction{FMain}{on\_write}
        \SetKwProg{Fn}{function}{}{}
        \Fn{\FMain{}}
        {
            \uIf{!handshake\_completed}{
                 \uIf{!$do\_handshake(data, data\_len$)} {
                    \KwRet $failure$\
                }
                \Else{
                    handshake\_completed = true
                } 
            }
            
            \While{true}{
                
                \uIf{send\_queue.size() $<=$ 0)}{
                    break \\
                }
                buf = send\_queue.front() \\
                pkt\_buf, error = \_conn\_write\_pkt(buf) \\
                \uIf{error != null} {
                    \KwRet $failure$\
                }
                
                \textcolor{red_color}{\texttt{write\_streams}(pkt\_buf)} \\
                \KwRet 
            }
        }
            
        \SetKwFunction{FMain}{write\_streams}
        \SetKwProg{Fn}{function}{}{}
        \Fn{\FMain{buf}}
        {
            \uIf{(\textcolor{blue_color}{this$\rightarrow$openflow)}}{
                \textcolor{blue_color}{int stream\_id = \texttt{generate\_stream\_id\_divisisble\_by\_3()}} \\
            }
            \uElseIf{(\textcolor{blue_color}{this$\rightarrow$ovsdb)}}{
                \textcolor{blue_color}{int stream\_id = \texttt{generate\_normal\_stream\_id()}} \\
            }
            \textcolor{red_color}{\texttt{on\_write\_stream(stream\_id, buf)}} \\
            \KwRet\
        }  
  
        \SetKwFunction{FMain}{on\_write\_stream}
        \SetKwProg{Fn}{function}{}{}
        \Fn{\FMain{stream\_id, buf}}
        {
            \While{true}{
                auto n = \texttt{\_conn\_write\_stream(ndatalen)} \\
                \uIf{n $>$ 0 \&\& and ndatalen > 0}{
                    \texttt{data.seek}(ndatalen) \\
                }
                \textcolor{red_color}{\texttt{send\_packet}()} \\
                \uIf{buf.size() = 0}{
                    $break$
                }
            }
            \KwRet\
        }  
        
        \SetKwFunction{FMain}{\textcolor{blue_color}{on\_ofl\_odb\_read}}
        \SetKwProg{Fn}{function}{}{}
        \Fn{\FMain{}}
        {
            \textcolor{blue_color}{array<uint8\_t, 65536> buf\_ofl}\\
            \textcolor{blue_color}{array<uint8\_t, 65536> buf\_odb} \\
            
            \uIf{\textcolor{blue_color}{activity detected of fp\_ofl}}
            {
                \uIf{(\textcolor{blue_color}{recvfrom(this$\rightarrow$fp\_ofl, buf\_ofl.data, buf\_ofl.len))}}{
                    \textcolor{blue_color}{send\_queue.push(buf\_ofl)}\\
                }
            }
            
            \uIf{\textcolor{blue_color}{activity detected of fp\_odb}}
            {
                \uIf{\textcolor{blue_color}{recvfrom(this$\rightarrow$fp\_odb, buf\_odb.data, buf\_odb.len))}}{
                    \textcolor{blue_color}{send\_queue.push(buf\_odb)}\\
                    \KwRet\
                }    
            }
        } 
\end{algorithm}

\subsection{QUIC Client and Server}
\gls{QUIC} uses a client-server model. 
The original development goal of \gls{QUIC} was to replace \gls{TCP} as a reliable transport protocol in HTTP3; however, the quicSDN architecture is different from HTTP3.
Specifically, unlike HTTP3, multiple applications interact with \gls{QUIC} in the quicSDN architecture, and these applications communicate with each end-point over the same \gls{UDP} connection.
In quicSDN's switch side implementation, \ovsdbserver{} and \ovsswitchd{} communicate with \quicclient{}.
On the controller side, \openflowryuapp{} and \ovsdbryuapp{} communicate with \quicserver{}.
We picked \texttt{\ngtcp{}} \cite{ngtcp2} for \gls{QUIC} code, because it is updated frequently with bug fixes. 
\texttt{\ngtcp{}} also provides easiness in feature extension in \gls{OVS} (as it is written in C) and is convenient for any feature integration with RYU code (via Python C extensions) on the controller side.

We divided \quicserver{} and \quicclient{} into server-agent, client-agent, and common APIs.
The server-agent APIs are responsible for serving requests from clients, invoking common APIs, negotiating versions, and completing \gls{QUIC} handshakes. 
The client-agent APIs are implemented to prepare the requests, rearrange the responses, and interact with common APIs. 
The common APIs are responsible for invoking the zero \gls{RTT} scenario, encrypting and decrypting packets, and storing the cryptographic keys.
This section presents the modifications relevant to quicSDN.

\subsubsection{QUIC Client}
\label{section_quic_client}

Algorithms \ref{alg:QUIC_CL_CODE} and \ref{alg:QUIC_Client_callbacks} present the pseudo-code of \quicclient{} module.
\quicclient{} spawns two \gls{UDP} servers listening on ports 6653 and 6640 on localhost (A\ref{alg:QUIC_CL_CODE}: L5-6)\footnote{This notation means Algorithm 1, Lines 5 through 6.} to intercept all connection requests and data packets from \ovsdbserver{} and \ovsswitchd{}. 
There are three sockets in \quicclient{}: two sockets for the above-mentioned \gls{UDP} servers, and one socket for connecting to \quicserver{}.
For these three sockets, we define three C structures to store connection information such as IP address, port, and socket information. 
These three C structures are for \ovsswitchd{} (\texttt{struct p\_openflow}), \ovsdbserver{} (\texttt{struct p\_ovsdb}), and \quicserver{} (\texttt{struct p\_quic}).

\texttt{\ngtcp{}} allows only one IP address and port to be specified in the \gls{CLI} commands.
We developed a new \gls{CLI} command to populate the above-mentioned three C structures (A\ref{alg:QUIC_CL_CODE}: L3) on the  \quicclient{} side:
\begin{itemize} \footnotesize
    \setlength{\itemindent}{-1em}
    \item  [\texttt{\$}] \texttt{<quic\_client\_path> <quic server addr> <quic server port> <openflow port> <ovsdb port>}
\end{itemize}
To support asynchronous I/O operations, each socket is mapped to a stream pointer \gls{FD}.
In conjunction with existing \gls{FD} (A\ref{alg:QUIC_CL_CODE}: L17) for a socket connected to \quicserver{}, two more \glspl{FD} named \texttt{fp\_ofl} (A\ref{alg:QUIC_CL_CODE}: L12-13) and \texttt{fp\_odb} (A\ref{alg:QUIC_CL_CODE}: L14-15) are introduced for each socket connected to \ovsswitchd{} and \ovsdbserver{}, respectively.
Any activity detected on these \glspl{FD} invokes a callback function.
We developed two new callback functions, \texttt{ofl\_cb()} and \texttt{odb\_cb()} (A\ref{alg:QUIC_CL_CODE}: L46-53) for \ovsdbserver{} and \ovsswitchd{} and modified the existing ones, \texttt{readcb()} (A\ref{alg:QUIC_CL_CODE}: line 40) and \texttt{writecb()} (A\ref{alg:QUIC_CL_CODE}: L43).
\texttt{readcb()} is invoked when the \gls{FD} receives a packet.
Inside \texttt{readcb()}, \texttt{on\_read()} receives data from the socket and passes it to \texttt{feed\_data()} (A\ref{alg:QUIC_Client_callbacks}: L8-20), which is responsible for processing \gls{QUIC} handshake and data packets.
If the \gls{QUIC} handshake has been completed successfully, then it is confirmed that all the necessary security keys are in place (A\ref{alg:QUIC_Client_callbacks}: L23-27).
The function \texttt{\_con\_recv()} checks if the received packet contains the long or short \gls{QUIC} header by inspecting the most significant bit of octet 0 (0x80) (A\ref{alg:common_apis}: L1-6). 
The long header is used for \gls{QUIC} version  \cite{QUIC_Version_negotiation} and 1-\gls{RTT} keys negotiations, and the short header is used for subsequent data communications.
\texttt{crypt\_quic\_message()} (\ref{cryptQUIC}) (A\ref{alg:common_apis}: L6) parses the packet and performs all the necessary \gls{QUIC} related operations such as encrypting packets, decrypting packets, and key management. 

\texttt{writecb()} is invoked to send the packet to \quicserver{}.
The \gls{QUIC} handshake is initiated by the \texttt{on\_write()} function (A\ref{alg:QUIC_Client_callbacks}: L23).
Inside \texttt{on\_write()}, \texttt{\_conn\_write\_pkt()} encrypts the packet (A\ref{alg:common_apis}: 10-12).
\texttt{write\_streams()} is then called to check if the packet is destined for \ovsswitchd{} or \ovsdbserver{} in order to generate appropriate stream IDs (A\ref{alg:QUIC_Client_callbacks}: L38-41).

Packets that are received on the sockets connected to \ovsdbserver{} and \ovsswitchd{} invoke \texttt{odb\_cb()} and \texttt{ofl\_cb()} callbacks (A\ref{alg:QUIC_CL_CODE}: L46-53), respectively.
Both callbacks invoke \texttt{on\_ofl\_odb\_read()}, where packets are pushed to the \texttt{send\_queue()} for \gls{QUIC} processing (A\ref{alg:QUIC_Client_callbacks}: L54-61).

\begin{algorithm}[!htb]
    
	\footnotesize
	\SetInd{0.9em}{0.7em}
	    \caption{Pseudo-code of {\quicserver{}}}
	\label{alg:QUIC_Server}
    \SetKwFunction{FMain}{main}
    \SetKwProg{Fn}{function}{}{}
    \Fn{\FMain{}}
    {
        \textcolor{blue_color}{p\_openflow, p\_ovsdb}, p\_quic = \{sock, port, addr\} \\
        \textcolor{red_color}{server\_arg = \{key, cert, p\_ovsdb, p\_openflow, p\_quic\}} \\
        
        \textcolor{red_color}{Populate server\_arg from CLI}

        \textcolor{blue_color}{p\_openflow = Connect to \openflowryuapp{} on 6653 port} \\
        \textcolor{blue_color}{p\_ovsdb = Connect to \ovsdbryuapp{} on 6640 port} \\
         
        \uIf {!(\textcolor{red_color}{start\_server(server\_arg)}} { 
             \KwRet;
        } 
        \textcolor{blue_color}{\texttt{close(p\_openflow, p\_ovsdb, p\_quic)]
        }}\\
        \KwRet\;
    }
         
        \SetKwFunction{FMain}{start\_server}
        \SetKwProg{Fn}{function}{}{}
        \Fn{\FMain{server\_arg}}
        {
            s1 = server\_arg$\rightarrow$p\_openflow$\rightarrow$sock \\
            s2 = server\_arg$\rightarrow$p\_ovsdb$\rightarrow$sock \\
            sock = create UDP server to accept \quicclient{} connections \\
            \textcolor{blue_color}{File *fp\_ofl =  \texttt{fileno(s1)}} \\ 
            \textcolor{blue_color}{File *fp\_odb =  \texttt{fileno(s2)}} \\
            \textcolor{red_color}{File *fd\_ = \texttt{fileno(sock)}} \\ 
            
            \texttt{// set event callbacks} \\
            fd\_ $\rightarrow$ \texttt{readcd(), writecb()} \\
            fp\_ofl $\rightarrow$ \texttt{ofl\_cb()} \\
            fp\_odb $\rightarrow$ \texttt{odb\_cb()} \\
          
            \texttt{ev\_run(ev\_d, 0)} \\
        }   
        
        \SetKwFunction{FMain}{readcb}
        \SetKwProg{Fn}{function}{}{}
        \Fn{\FMain{ev\_loop *loop, ev\_io *w}}
        {
            auto c = <client *>w$\rightarrow$data \\
            \textcolor{red_color}{\texttt{on\_read()}} \\
        }  
        
        \SetKwFunction{FMain}{writecb}
        \SetKwProg{Fn}{function}{}{}
        \Fn{\FMain{ev\_loop *loop, ev\_io *w}}
        {
            auto c = <client *>w$\rightarrow$data \\
            \textcolor{red_color}{\texttt{on\_write()}} \\
        }  
        \SetKwFunction{FMain}{\textcolor{blue_color}{ofl\_cb}}
        \SetKwProg{Fn}{function}{}{}
        \Fn{\FMain{ev\_loop *loop, ev\_io *w}}
        {
            \textcolor{blue_color}{auto c = <ofl *>w$\rightarrow$data}\\
            \textcolor{blue_color}{this$\rightarrow$type\_flag = openflow} \\
            \textcolor{blue_color}{\texttt{on\_ofl\_odb\_read()}} \\
        } 
        
        \SetKwFunction{FMain}{\textcolor{blue_color}{odb\_cb}}
        \SetKwProg{Fn}{function}{}{}
        \Fn{\FMain{ev\_loop *loop, ev\_io *w}}
        {
           \textcolor{blue_color}{ auto c = <ofdb *>w$\rightarrow$data}\\
            \textcolor{blue_color}{this$\rightarrow$type\_flag = ovsdb} \\
            \textcolor{blue_color}{\texttt{on\_ofl\_odb\_read()}} \\
        } 
\end{algorithm}

\begin{algorithm}[!htb]
    
	\footnotesize
	\SetInd{0.9em}{0.7em}
	    \caption{Pseudo-code of {\quicserver{} (continued from Algorithm \ref{alg:QUIC_Server})}}
	\label{alg:QUIC_SERVER_CALLBACKS}
        
        \SetKwFunction{FMain}{on\_read}
        \SetKwProg{Fn}{function}{}{}
        \Fn{\FMain{}}
        {
            buffer<uint8\_t, int, port> buf \\
            \While{true}{
                \uIf{!(recvfrom(this$\rightarrow$fd\_, buf.data, buf.len))}{
                        return
                }
                
                hd = this$\rightarrow$hd \\
                \uIf {(buf[0] \& 0x80)} {
                    \texttt{\_pkt\_decode\_hd\_long(\&hd, \texttt{buf.data()}) }\\
                } \Else {
                    \texttt{\_pkt\_decode\_hd\_short(\&hd, \texttt{buf.data()})} \\
                }
            
                \textcolor{red_color}{\texttt{\_accept}(\texttt{buf.data(), buf.len)}} \\
            }
            
            \KwRet\
        }  
        
         \SetKwFunction{FMain}{on\_write}
        \SetKwProg{Fn}{function}{}{}
        \Fn{\FMain{}}
        { 
            \uIf{!(handshake\_completed)}{
                \texttt{do\_handshake()} \\
                handshake\_completed = true \\
            }
            \Else{
                \uIf{!schedule\_restransmit()}{
                    \KwRet $failure$\
                }
            }
            
            \textcolor{Mahogany}{buf = send\_queue.front()} \\
            \textcolor{blue_color}{stream\_id = \_conn\_map[buf$\rightarrow$port]} \\
            
            \texttt{on\_write\_stream(stream\_id)}\\
            
           % \EndFor
           
           \For{;;}{
                n = \textcolor{red_color}{\texttt{\_conn\_write\_pkt()}} \\
                \uIf{n = 0}{
                    $break$ \\
                }
                \texttt{buf\_.push(n);}. \\
                \textcolor{blue_color}{\texttt{send\_packet(buf)}} \\
           }
           
           \KwRet\
        }
        
        \SetKwFunction{FMain}{\textcolor{blue_color}{on\_ofl\_odb\_read}}
        \SetKwProg{Fn}{function}{}{}
        \Fn{\FMain{}}
        { 
             \While{true}{
                %\textbf{\textcolor{blue}{
               \textcolor{blue_color}{ buffer<uint8\_t, int, port> buf\_ofl} \\
                \uIf{(\textcolor{blue_color}{recvfrom(this$\rightarrow$fp\_ofl, data, datalen))}}{
                        %\textbf{\textcolor{blue}{
                        \textcolor{blue_color}{buf\_ofl.data = data} \\ 
                        %\textbf{\textcolor{blue}{
                        \textcolor{blue_color}{buf\_ofl.len = datalen} \\
                        %\textbf{\textcolor{blue}{
                        \textcolor{blue_color}{buf\_ofl.port = this$\rightarrow$port;}\\
                        \tcp{e.g 6653 for openflow}
            
                        \textcolor{red_color}{\texttt{send\_queue.push(buf\_ofl)}}\\
                }
            
                \textcolor{blue_color}{buffer<uint8\_t, int, port> buf\_odb;} \\
                \uIf{(\textcolor{blue_color}{recvfrom(this$\rightarrow$fp\_odb, data, datalen))}}{
                        %\textbf{\textcolor{blue}{
                        \textcolor{blue_color}{buf\_odb.data = data}\\ 
                        %\textbf{\textcolor{blue}{
                        \textcolor{blue_color}{buf\_odb.len = datalen} \\
                        %\textbf{\textcolor{blue}{
                       \textcolor{blue_color}{ buf\_odb.port = this$\rightarrow$port} \\
                        \tcp{e.g 6640 for ovsdb}
                
                        \textcolor{red_color}{\texttt{send\_queue.push(buf\_odb)}} \\
                        %return;
                }
            }
           
           \KwRet\
        }
        
\end{algorithm}

\begin{algorithm}[!htb]
    
	\footnotesize
	\SetInd{0.9em}{0.7em}
	    \caption{\textbf{Common APIs}}
	\label{alg:common_apis}

        \SetKwFunction{FMain}{\_con\_recv}
        \SetKwProg{Fn}{function}{}{}
        \Fn{\FMain{data, datalen, \&s}}
        {
            \uIf{(data[0] \& 0x80)}{
                \texttt{\_pkt\_decode\_hd\_long()} \\
            }
            \Else{
                \texttt{\_pkt\_decode\_hd\_short()} \\
            }
            
            \texttt{crypt\_quic\_message(decrypt)} \\
        }

        \SetKwFunction{FMain}{\_conn\_write\_stream}
        \SetKwProg{Fn}{function}{}{}
        \Fn{\FMain{}}
        {
            
            \texttt{find\_stream\_info(dest)} \\
            \texttt{\_conn\_write\_pkt()}
        }
        
        \SetKwFunction{FMain}{\_conn\_write\_pkt}
        \SetKwProg{Fn}{function}{}{}
        \Fn{\FMain{}}
        {
            \texttt{conn\_write\_probe\_pkt()} \\
            \texttt{crypt\_quic\_message(encrypt)} \\
        }
        
        \SetKwFunction{FMain}{\_accept}
        \SetKwProg{Fn}{function}{}{}
        \Fn{\FMain{data, datalen}}
        {
            plain\_text = \texttt{crypt\_quic\_message(decrypt)} \\
            \uIf{(\textcolor{blue_color}{this$\rightarrow$stream\_id is divisible by 3)}}{
                \textcolor{blue_color}{\_conn\_map[opnflow\_port] = stream\_id}
                \textcolor{blue_color}{\texttt{sendto(this$\rightarrow$fp\_ofl)}}
            }
            \Else{
                \textcolor{blue_color}{\_conn\_map[odb\_port] = stream\_id}
                \textcolor{blue_color}{\texttt{sendto(this$\rightarrow$fp\_odb)}}
            }
            
        }
    
\end{algorithm}

\subsubsection{QUIC Server}
\label{section_quic_server}

Algorithms \ref{alg:QUIC_Server} and \ref{alg:QUIC_SERVER_CALLBACKS} present the \quicserver{} implementation.
\quicserver{} connects to \openflowryuapp{} and \ovsdbryuapp{} modules (A\ref{alg:QUIC_Server}: L5-6), which are listening on ports 6653 and 6640, respectively.
Similar to \quicclient{}, there are three sockets in \quicserver{}. Two sockets are for \openflowryuapp{} and \ovsdbryuapp{} for port 6653 and 6640, respectively, while one socket is for a connection to  \quicclient{}.
In order to store the connection information of these three sockets, we define three C structures, \texttt{p\_quic}, \texttt{p\_openflow} and \texttt{p\_ovsdb} for \quicclient{}, \openflowryuapp{}, and \ovsdbryuapp{} respectively.
%Similar to \quicclient{}, we define three C structs, \texttt{p\_quic}, \texttt{p\_openflow} and \texttt{p\_ovsdb}, for \quicserver{}, \openflowryuapp{}, and \ovsdbryuapp{} (A\ref{alg:QUIC_Server}: L2).
%\texttt{p\_openflow} and \texttt{p\_ovsdb} store the connection information for northbound connections to \openflowryuapp{} and \ovsdbryuapp{}, while \texttt{p\_quic} stores the connection information for the southbound connection to \quicclient{}. 
As previously mentioned, \ngtcp{}'s \gls{CLI} commands contain only one IP address and port; therefore, we developed a new \gls{CLI} command for \quicserver{} to populate the three C structs with the appropriate information (A\ref{alg:QUIC_Server}: L3):
\begin{itemize} \footnotesize
    \setlength{\itemindent}{-1em}
    \item  [\texttt{\$}] \texttt{<quic\_server\_path> <quic server addr> <quic server port> <key> <certificate> <openflow port> <ovsdb port>}
\end{itemize}
The above-mentioned three sockets in \quicserver{} are capable of asynchronous I/O operations (A\ref{alg:QUIC_Server}: L14).
%On the \quicserver{}, above-mentioned three sockets are capable of asynchronous I/O operations (A\ref{alg:QUIC_Server}: L14).
Three \glspl{FD} are mapped as stream pointers to these three sockets.
Among the three \glspl{FD}, the existing \gls{FD} (\texttt{fd\_}) is modified, and two new \glspl{FD}, \texttt{fp\_ofl} and \texttt{fp\_odb} (A:\ref{alg:QUIC_Server} L15-16), are added.
The \gls{FD} \texttt{fd\_} is for the \gls{QUIC} Connection to \quicclient{}, \gls{FD} \texttt{fp\_ofl} is for the connection to \openflowryuapp{}, and \gls{FD} \texttt{fp\_odb} is for the connection to \ovsdbryuapp{}.
These \glspl{FD} monitor the sockets via an event loop and invoke callbacks if any activity is detected.
Callbacks \texttt{readcb()} and \texttt{writecb()} are invoked if activity is detected on \texttt{fd\_}.
Similarly, callback \texttt{ofl\_cb()} (A:\ref{alg:QUIC_Server} L29-32) is invoked if any activity is detected on \texttt{fp\_ofl}, and callback \texttt{odb\_cb()} (A:\ref{alg:QUIC_Server} L33-36) is invoked if any activity is detected on \texttt{fp\_odb}.
%Algorithm \ref{alg:QUIC_SERVER_CALLBACKS} describes the callbacks. 
%\texttt{readcb} invokes \texttt{on\_read} and \texttt{writecb} invokes \texttt{on\_write}  for reading and writing to the \gls{FD}. 

The \texttt{on\_read()} function is responsible for reading \texttt{fd\_} to process the received \gls{QUIC} packets (A:\ref{alg:QUIC_SERVER_CALLBACKS} L1-12).
First, the received \gls{QUIC} packet is evaluated to check if the header in the corresponding packet is a long header or a short header (A:\ref{alg:QUIC_SERVER_CALLBACKS} L7). 
Then the packet is passed to the \texttt{\_accept()} function for decryption (A:\ref{alg:QUIC_SERVER_CALLBACKS} L11). 

The \texttt{on\_write()} function is for sending packets to \quicclient{} (A:\ref{alg:QUIC_SERVER_CALLBACKS} L13-29). 
This function first evaluates and performs a \gls{QUIC} handshake with \quicclient{} to exchange cryptographic keys (A:\ref{alg:QUIC_SERVER_CALLBACKS} L15). 
The buffer received in \texttt{on\_write()} contains the OpenFlow or \gls{OVSDB} port information to maintain an external map (\texttt{\_conn\_map}) of port to \texttt{streamID} mapping for reverse traffic (A:\ref{alg:QUIC_SERVER_CALLBACKS} L21). 
The function \texttt{on\_write\_stream()} searches for an existing stream, and if the stream does not exist yet, it opens a new stream and packs the data into it (A:\ref{alg:QUIC_SERVER_CALLBACKS} L22). 
\texttt{\_conn\_write\_pkt()} is responsible for encrypting the packets and placing them into the transmission queue (A:\ref{alg:common_apis} L10-12).

The function \texttt{on\_ofl\_odb\_read()} is called by \texttt{ofl\_cb()} and \texttt{odb\_cb()} callbacks (A:\ref{alg:QUIC_SERVER_CALLBACKS} L30-44). 
This function is responsible for exchanging packets between \openflowryuapp{}, \ovsdbryuapp{}, and \quicserver{}.

\subsubsection{\texttt{crypt\_quic\_message()}} 
\label{cryptQUIC}
This \gls{API} consists of the \texttt{decrypts\_message()} and \texttt{encrypts\_message()} functions, responsible for decrypting and encrypting packets in multiple phases.
Each phase has a different set of keys.
Before acquiring the symmetric keys, \gls{QUIC} completes four phases.
The first phase is the Initial Key Agreement, where each party sets and exchanges the initial key and additional information, such as HMAC.
Both parties then agree to a common key ($ik$), which is derived from the Client Initial Key ($ik_c$) and the Server Initial Key ($ik_s$). 
The second stage is the Initial Data Exchange, where data is encrypted and decrypted by using an \gls{AEAD} Scheme \cite{rogaway2002authenticated} and $ik$. 
The third phase is the Key Agreement, where the session key ($k$) is derived from the client session key ($k_c$), server session key ($k_s$), and $aux$, where $aux \in \{0,1\}$.
The fourth phase is the Data Exchange. 
In this phase, data is sent using the associated \gls{AEAD} scheme and $k$. 
The server uses $k_c$ to encrypt and $k_s$ to decrypt packets, while the client uses $k_s$ to encrypt and $k_c$ to decrypt packets. 
In addition, \texttt{crypt\_quic\_message()} also prepares the initialization vector ($iv$) and salt for the cryptographic keys.

\subsection{RYU}
In tcpSDN, RYU inherits the \gls{TCP} transport layer infrastructure in the form of base classes. 
The most important base class is \texttt{OFPHandler}, which declares a controller base class object called \texttt{OpenFlowController}. 
Inside this object, a server is spawned to receive and process all the received packets via an event loop.
%The unmodified \gls{TCP}-based RYU architecture spawns a server, which is tied to an event loop in \texttt{OpenFlowController} via the \texttt{recv\_loop} function. 
Any packet received will be pushed to the RYU app for processing. 
In quicSDN, to make RYU compatible with \gls{UDP}, the first task is to replace the \gls{TCP} infrastructure and have RYU spawn a \gls{UDP} server instead. 
This modification is challenging due to RYU's current event loop callback mechanism.
This callback mechanism is based on asynchronous \gls{TCP} socket, which is a part of eventlet I/O library.
In order to make it \gls{UDP} based, first the eventlet library needs to be changed to support \gls{UDP} sockets. 
Moreover, RYU needs to implement the server based on the eventlet library \gls{UDP} sockets.
%and accordingly process the \gls{UDP} packets before pushing them to the RYU. 
%in the mechanism RYU currently have for callbacks, under the event loop. 
In quicSDN, the RYU applications \openflowryuapp{} and \ovsdbryuapp{} are started using the following \glspl{CLI} commands:
\begin{itemize} \footnotesize
    \setlength{\itemindent}{-1em}
    \item  [\texttt{\$}] \texttt{ryu-manager --ofp-listen-port <port num> <app name>}
    \item [\texttt{\$}]
    \texttt{ryu-manager <OVSDB app name>}
\end{itemize}
Note that no changes were made to the state machines of \openflowryuapp{} and \ovsdbryuapp{}.
In tcpSDN, packets are processed in the eventlet library which is implemented using GreenSocket. 
We modified the RYU event loop to make sure that packets are directly pushed from eventlet library to be processed in the app itself.
The OpenFlow (\openflowryuapp{}) and \gls{OVSDB} (\ovsdbryuapp{}) controllers both use the same transport layer infrastructure.

\begin{figure}[t]
    \centering
    \includegraphics[width=1.0
    \linewidth]{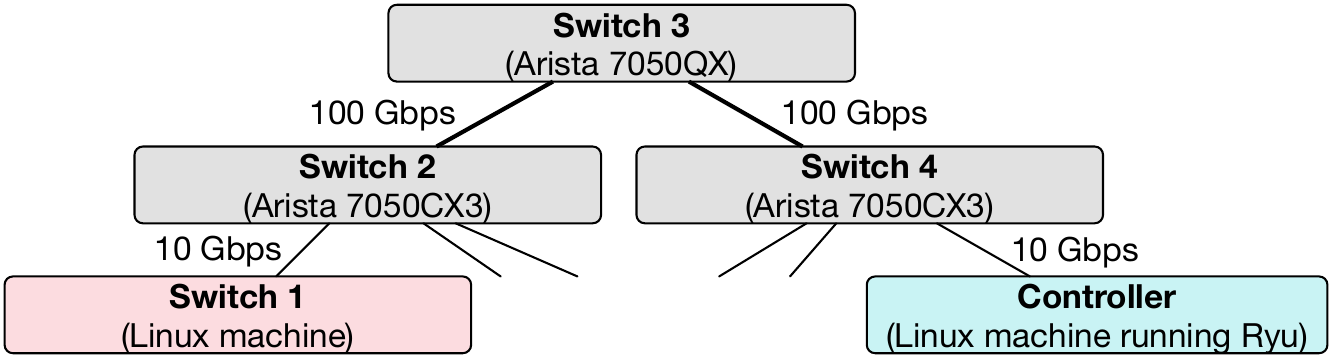}
    \captionsetup{font=footnotesize}
    \caption{Testbed topology. Switch 1 and the Controller communicate through Switch 2, 3, and 4.
    }
    \label{fig:topologyModeling}
\end{figure}
\begin{figure}[t]
    \centering
    \includegraphics[width=1.\linewidth]{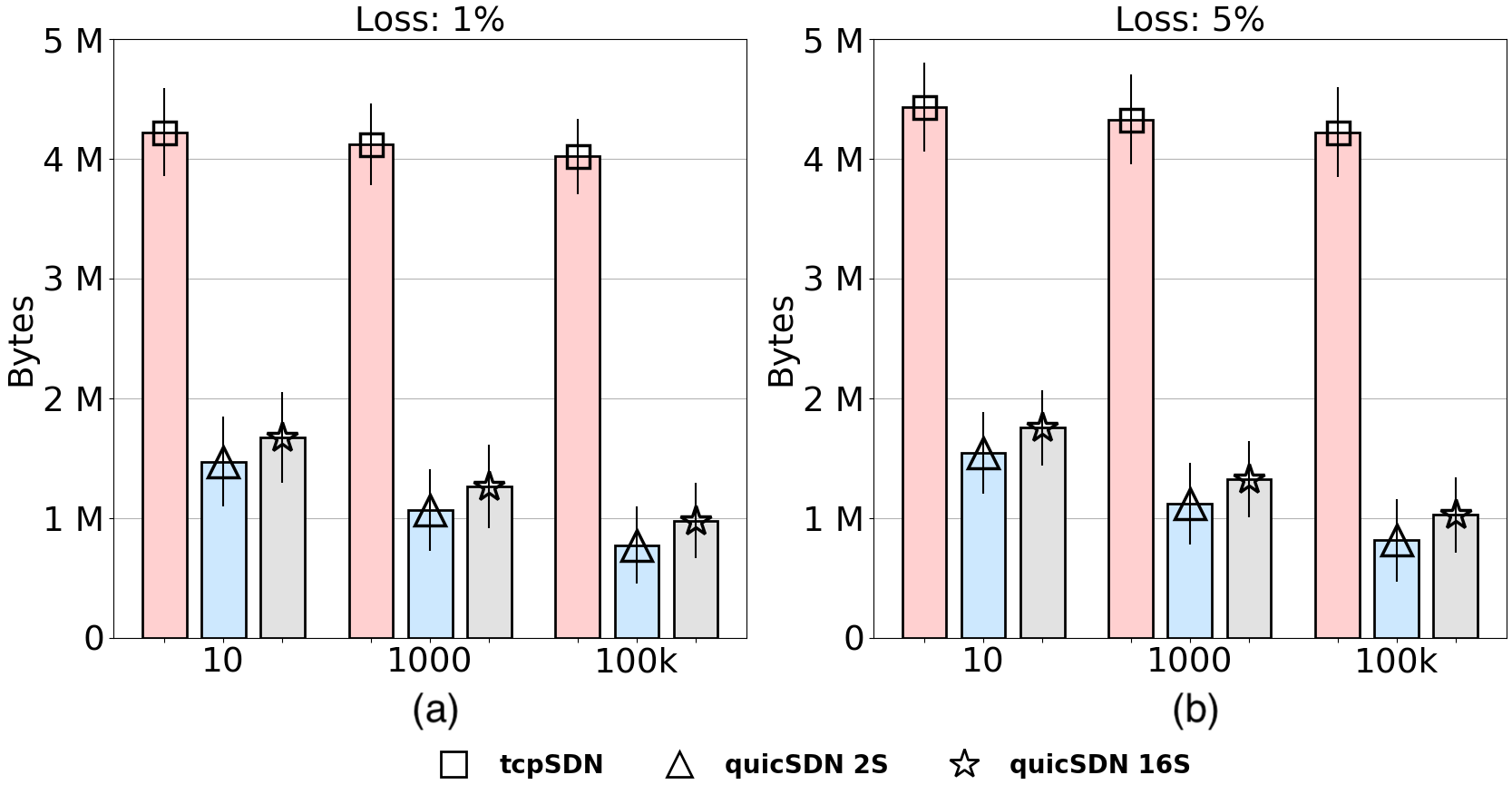}
    \captionsetup{font=footnotesize}
    \caption{The protocol overhead of tcpSDN and quicSDN in the \textit{short \gls{RTT}} scenario. 
    The x-axis shows three message generation rates: 10, 1000, 100k messages per sec. 
    Sub-figures (a) and (b) present the overhead for 1\% and 5\% packet loss rates, respectively.
    }
    \label{fig:overheadWithLossMI}
\end{figure}

\section{Empirical Evaluation}
\label{emp_eval}
In this section, we empirically evaluate quicSDN versus tcpSDN.
The testbed configuration is demonstrated in Figure \ref{fig:topologyModeling}.
% which is similar to the leaf-spine topology.
We use three high-performance switches: one 7050QX (Switch 3) and two 7050CX3 (Switch 2 and 4).
For Switch 1, we use a Linux machine, which allows us to run OpenFlow and OVSDB, as well as emulating the execution of more agents on the switch to generate various traffic patterns.
% We connected the Controller to Switch 4 instead of Switch 3 to increase the communication path 
The \gls{RTT} between Switch 1 and the Controller shows a mean value of 0.32 ms and a standard deviation of 0.051 ms; we refer to this scenario as \textit{short RTT}.
\begin{figure}[t]
    \centering
    \includegraphics[width=1.\linewidth]{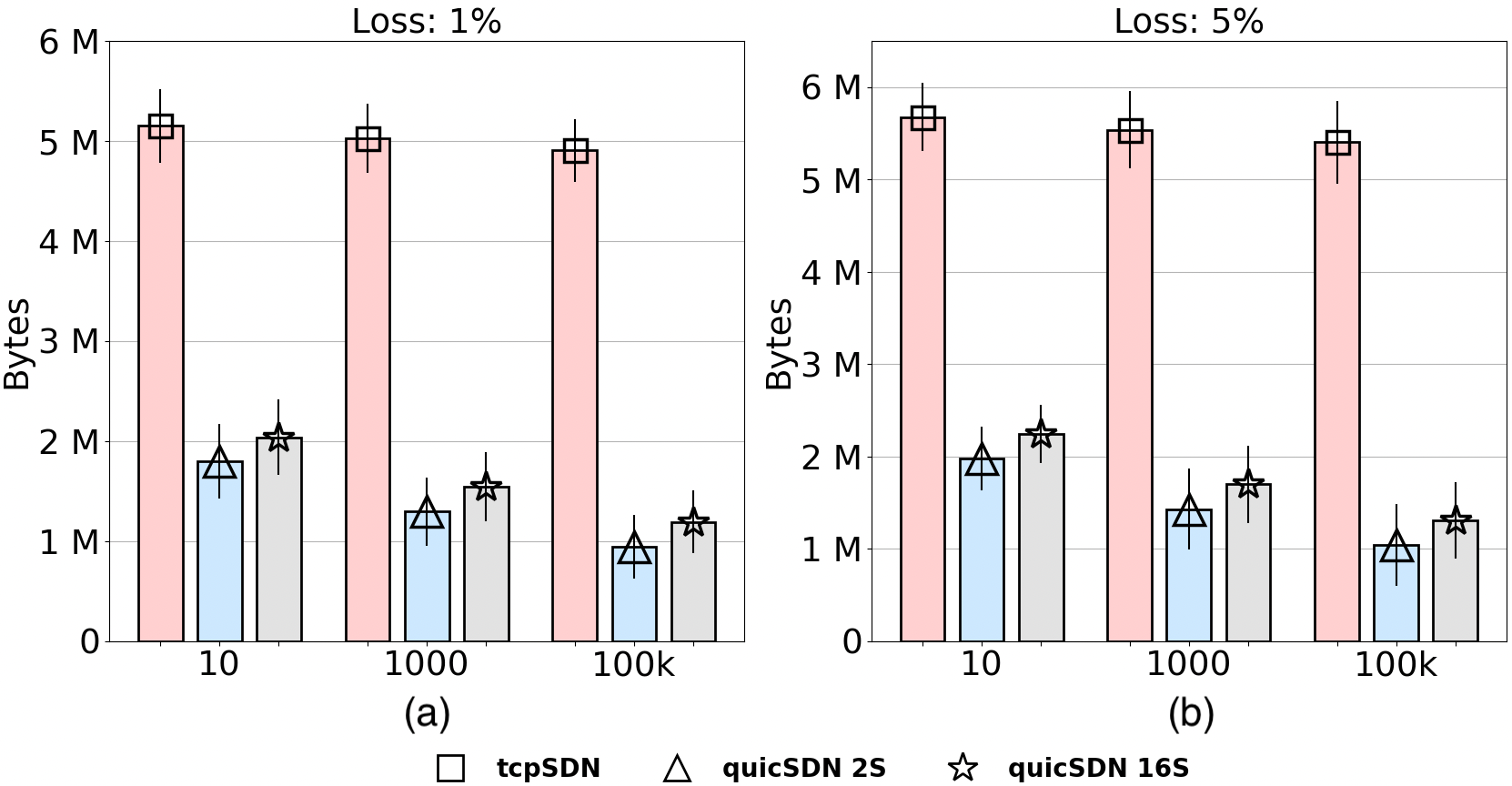}
    \captionsetup{font=footnotesize}
    \caption{The protocol overhead of tcpSDN and quicSDN in the \textit{long \gls{RTT}} scenario. 
    The x-axis shows three message generation rates: 10, 1000, 100k messages per sec. 
    Sub-figures (a) and (b) present the overhead for 1\% and 5\% packet loss rates, respectively.
    }
    \label{fig:delayRTTOverhead}
\end{figure}
We also configured the testbed to generate considerably higher RTT values with a mean value of 150 ms and a standard deviation of 50 ms; we refer to this scenario as \textit{long RTT}.
This long \gls{RTT} represents a scenario where the data plane is controlled by a remote controller or when there is a considerable traffic competing with southbound control traffic.
For each experiment performed, the total number of messages generated from Switch 1 to the Controller and vice-versa are 50,000 on each side. 
Also, we vary the message generation rate per seconds as 10, 1000, 10,000 messages per second.
% For each message received on the controller, a reply is sent back to the switch as well.
To represent various congestion levels in the switches, we introduce packet loss rates of 1\% and 5\%.
\gls{TSO} and \gls{GRO} have been disabled on Switch 1 and the Controller.

We consider two variants of quicSDN and compare them against tcpSDN.
These variants are explained as follows. 
\textbf{quicSDN-2s}: In this implementation, all the messages generated by OpenFlow are assigned to stream 0 and all the OVSDB generated messages are assigned to stream 2.
Note that these streams are bidirectional; for example, stream ID 0 is used for bidirectional OpenFlow messages between Switch 1 and the Controller.
\textbf{quicSDN-16s}: This implementation opens 16 bidirectional streams. 
As explained in Section \ref{streamID}, each messages is assigned a different stream identifier in the range 0 to 15, in a round-robin fashion.
% We use 16 stream IDs in our implementation.

\subsection{Overhead}
\label{empirical_overhead_eval}
We first evaluate the communication overhead between Switch 1 and the Controller for exchanging OpenFlow and OVSDB messages.
Whenever a controller sends a message (command) to the switch, the Controller needs to receive a reply from the switch to ensure proper enforcement of the commands.
Therefore, we measure the \textit{bidirectional} communication overhead between the two nodes.
Figures \ref{fig:overheadWithLossMI} and \ref{fig:delayRTTOverhead} present protocol overhead for the short-\gls{RTT} and long-\gls{RTT} scenarios, respectively.
The results show that the overhead of quicSDN is considerably lower than tcpSDN.
As we increase the number of messages exchanged per second between the switch and the controller, the communication overhead of quicSDN and tcpSDN drops; also, the decline of quicSDN overhead is higher than tcpSDN.
These empirical results conform to the analytical results we presented in Section \ref{theor_comm_overhead}.
The lower overhead of quicSDN is due to two main reasons:
\textit{First}, multiple agents (OpenFlow and OVSDB in these experiments) use a single connection to communicate with the controller. 
Therefore, by multiplexing the messages of these two agents, quicSDN achieves a lower probability of sending packets smaller than \gls{MTU}, compared to tcpSDN.
% Also, as it can be observed, increasing message rate results in more reduction in overhead when using quicSDN.
\textit{Second}, the transport header overhead of quicSDN is lower than tcpSDN when the number of streams is less than ten.
As we increase the number of streams from two to ten, the number of QUIC frames per packet may increase as well, depending on the message size.
Although a higher number of streams results in a better mitigation of the \gls{HOL} problem, this benefit comes at the cost of higher transport layer overhead. 
Nevertheless, as the results show, the overhead of quicSDN-16s is still lower than tcpSDN.

Figures \ref{fig:overheadWithLossMI} and \ref{fig:delayRTTOverhead} show that the overhead of tcpSDN and quicSDN are higher in the long RTT scenario compared to the short RTT scenario.
This overhead is caused by the fluctuations in \gls{RTT}, which cause out of order packets and more number of retransmissions.
However, the overhead increase of quicSDN is lower than that of tcpSDN.
For example, for 1\% loss rate and 10 messages/second, increasing the RTT causes the overhead of tcpSDN to increase by 25.9\%, while the overhead of quicSDN-2s increases by 19.5\%.
% There are two reasons for quicSDNs' better performance in overhead reduction.
% First reason is 
% The \gls{QUIC}'s default capability of reporting \textbf{\gls{NACK}}.
%\gls{NACK} is a mechanism in which a receiver can notify sender about the packets it has not received. 
\gls{TCP} does not support \gls{NACK}, instead it supports \gls{SACK}\footnote{This feature must be enabled as a socket option.}.
While \gls{NACK} specifies the packets that a receiver has \textit{not} received, \gls{SACK} specifies the range of packets a receiver has received.
These notifications are specified as ranges.
\gls{TCP} allows including up to three \gls{SACK} blocks  \cite{mathis1996rfc2018}, whereas, \gls{QUIC} supports 256 \gls{NACK}  blocks \cite{newQUICRFC}. 
This makes a difference when packet loss is eminent, as supporting only three blocks cause more number of ACKs with \gls{SACK} options than the number of ACKs with \gls{NACK}.

% The overhead from range of packets size perspective is shown in Figure \ref{fig:Ecdf}.
% As figure shows, tcpSDN has more packets of smaller sizes in all scenarios.
% This is because in tcpSDN, irrespective of their payload size, each message bears the cost of the \gls{TCP} header.
% Therefore, the protocol overhead for tcpSDN is higher than quicSDN.

%
% \begin{figure}[t]
%     \centering
%     \includegraphics[width=1.0\linewidth]{TopoMessageDelivery1.pdf}
%     \captionsetup{font=footnotesize}
%     \caption{Figure (a) shows the topology used in the experiment to determine the message delivery delay. Both switch and controller are on the same machine to avoid any ethernet synchronous timing mismatch. Message delivery delay is calculated by determining the time elapsed from sending a message and receiving it. Figure (b) shows the mechanism of constructing a \gls{QUIC} message from segments of a large message combining with smaller messages. 
%     }
%     \label{fig:messageDeliveryDelayTopo}
% \end{figure}

\subsection{Message Delivery Delay}
\label{message_delivery_delay}
In this experiment, we measure message delivery delay between the Controller and Switch 1 (in Figure \ref{fig:topologyModeling}).
We define message delivery delay as the time interval between the time instance a message is generated by an application running on the sender until complete message reception by the application running on the receiver.
We use two message size pairs: <400,600>, and <3000,600>.
Each pair represents the size of messages generated by the sender's application.
We also vary \gls{RTT} and packet loss rate between the two nodes.

\begin{figure*}[t]
    \centering
    \includegraphics[width=1.0\linewidth]{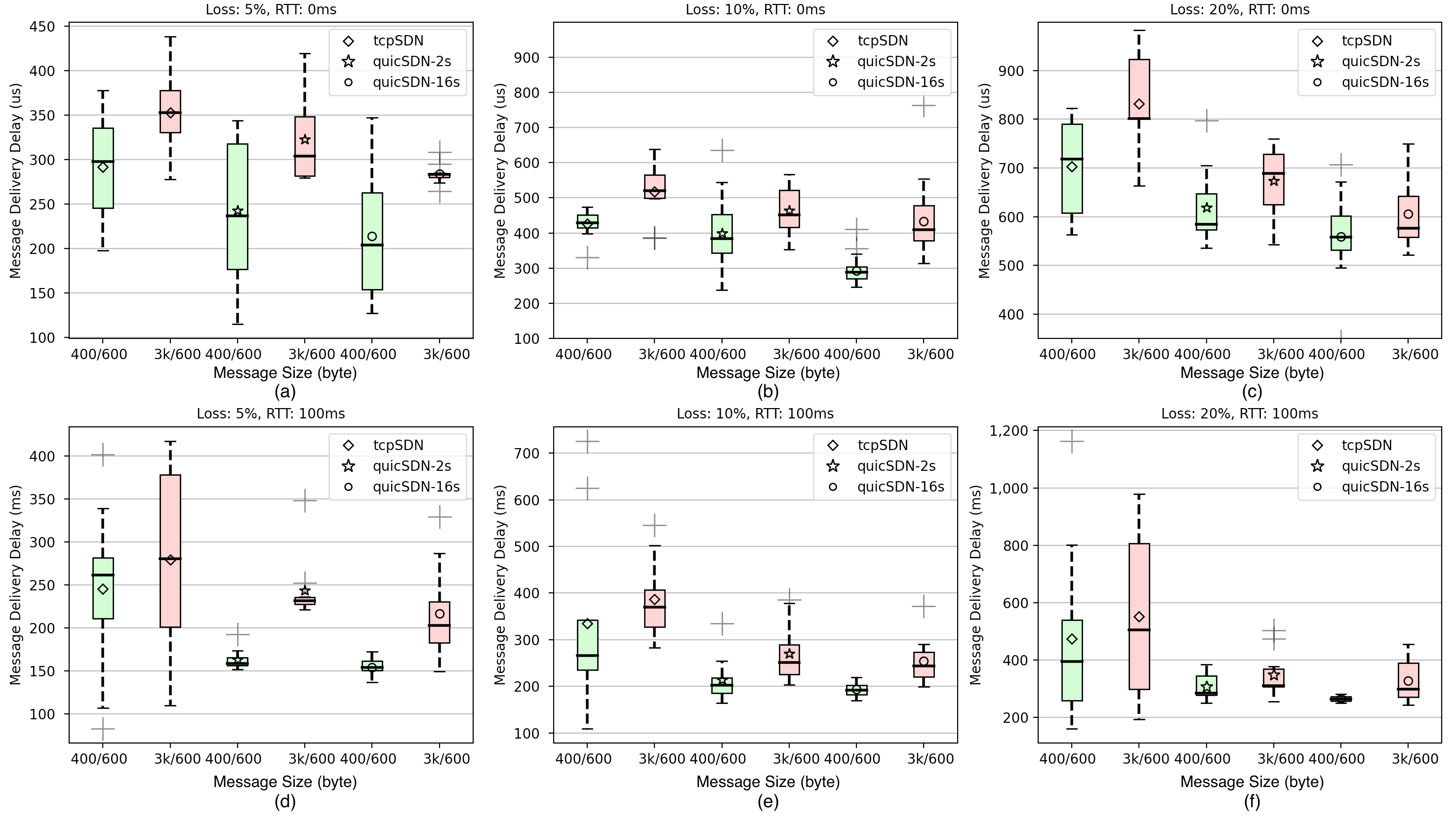}
    \captionsetup{font=footnotesize}
    \caption{Message delivery delay.
    The experiment is performed 30 times for each configuration.
    The results confirm the lower message delivery delay of quicSDN compared to tcpSDN. Also, quicSDN-16s provides shorter delivery delay compared to quicSDN-2s because using a higher number of streams increases the effectiveness of determining independence among packet frames.
    }
    \label{fig:messageDeliveryDelayGraph}
\end{figure*}

Figure \ref{fig:messageDeliveryDelayGraph} shows the result.
First, we observe that the message delivery delay of tcpSDN is higher than quicSDN-2s and quicSDN-16s.
The primary reason behind this enhancement is that, in contrast with tcpSDN, quicSDN can immediately deliver a stream payload to the application if it contains an entire message, regardless of the loss of packets with smaller sequence number.
Second, the message delivery delay of quicSDN-2s is higher than quicSDN-16s.
This is because, as the number of streams increases, QUIC can determine the independence between frames (inside packets) more effectively.
The final observation is that for both quicSDN-2s and quicSDN-16s, message delivery delay is more fluctuating than tcpSDN, and this is due to the order of packet arrivals and interdependence between frames included in each packet.
In the following, we present more details regarding the underlying causes of the higher performance of quicSDN compared to tcpSDN.

\begin{figure}[t]
    \centering
    \includegraphics[width=1.0\linewidth]{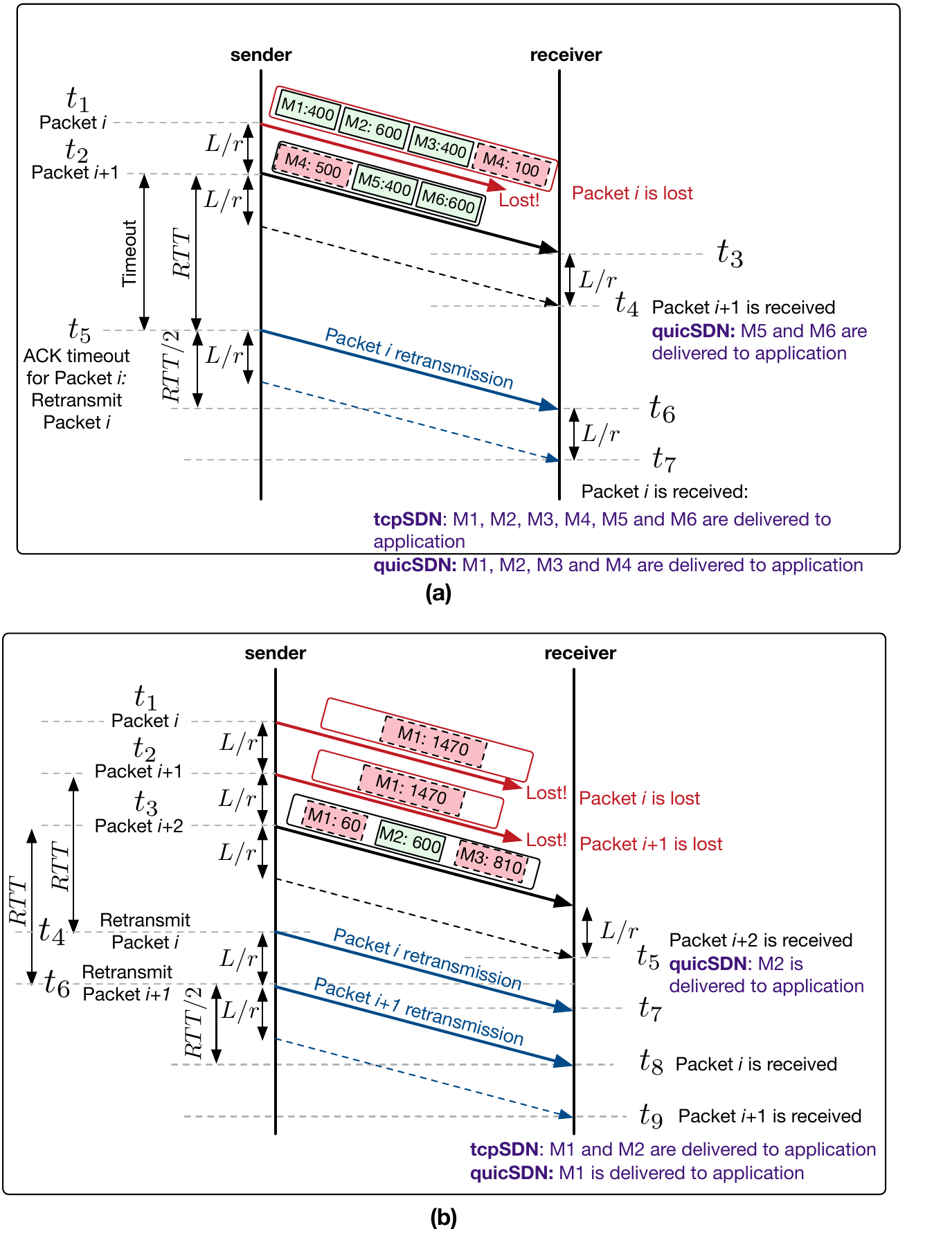}
    \captionsetup{font=footnotesize}
    \caption{The effect of packet loss and message size on the message delivery delay of tcpSDN and quicSDN.
    Here, $L$ is the packet length (bits) and $r$ is the transmission rate (bps) between the client and server.
    Red boxes show partial message inclusion in a packet and green boxes show complete inclusion of a message in a packet.
    }
    \label{fig:messageDeliveryDelayDiag}
\end{figure}

Assume an application generates message sizes 400 and 600 bytes, simply denoted as <400,600>.
Figure \ref{fig:messageDeliveryDelayDiag}a presents a sample packet transmission from a sender to a receiver.
Assume each packet can include 1470 bytes of message data.
At time $t_{1}$, the sender generates and includes four messages in Packet $i$: M1 (400 bytes), M2 (600 bytes), M3 (400 bytes), and part of M4 (600 bytes).
Packet $i$ includes 100 bytes of M4, and the remaining 500 bytes of this message are included in Packet $i+1$.
Packet $i$, which is transmitted at time $t_{1}$, is lost along the path between the sender and receiver.
Packet $i+1$ transmitted at time $t_2$ is fully received at $t_4$.
Note the transmission time of each packet is $L/r$, where $L$ is packet size and $r$ is the link speed between the two nodes. Propagation delay between the two nodes is $RTT/2$.
At $t_4$, in a quicSDN network, since messages M5 and M6 are fully received, they are delivered to the application, without having to wait for Packet $i$.
In contrast, in a tcpSDN network, since the receiver needs to receive all the bytes in order, it needs to wait for Packet $i$ before processing Packet $i+1$.
Assuming the \gls{RTO} of the sender is $RTT$, Packet $i$ is retransmitted at $t_5$.\footnote{\gls{RTO} value is usually higher than $RTT$ to account for RTT variations.}
% and delivered in $RTT + RTT/2 + 2L/r$ to the receiver after its initial transmission time $t_{1}$.
At $t_7$, Packet $i$ is fully received and tcpSDN delivers all the sent messages to the application, while quicSDN delivers the remaining messages only, i.e., M1, M2, M3, and M4.
In this example, for quicSDN, the delivery delay of M5 and M6 is $RTT/2 + L/r$.
With tcpSDN, the delivery delay of these two messages is $RTT + RTT/2 + 2L/r$.
In a realistic scenario, for example, if M5 and M6 are two flow rules, the switch can process and install these rules with a shorter delay compared to tcpSDN, thereby resulting in faster reaction to network dynamics.

Figure \ref{fig:messageDeliveryDelayDiag}b presents a sample packet transmission from a sender to a receiver where the message sizes generated by the sender are <3000,600>.
At time $t_{1}$, the sender includes the constituting bytes of message M1 (3000 bytes) in Packet $i$, Packet $i+1$, and Packet $i+2$.
Packet $i+2$ includes the remaining bytes of messages M1 (60 bytes), message M2 (600 bytes), and some bytes of message M3 (810 bytes).
Packets $i+1$ and $i+2$ transmitted at $t_1$ and $t_2$ are lost.
At time $t_5$, packet $i+1$ is fully received.
At this time, in a quicSDN network, M2 is delivered to the application. 
% therefore, the delivery delay of this message is $RTT/2 + L/r$.
In a tcpSDN network, since the delivery of M2 is contingent upon the successful reception of all these three packets, messages M1, M2 and M3 are delivered to the application at time $t_9$.
% Therefore, the delivery delay of M2 when using tcpSDN is 
This scenario justifies the effect of increasing message size on message delivery delay and conforms with the empirical results presented in Figure \ref{fig:messageDeliveryDelayGraph} that show delivery delay is higher for larger message sizes.

% REMOVED NAGLE'S ALGORITHM EFFECT
% In addition to the above mentioned justification, packet delivery of tcpSDN is affected by the following.
% %
% In tcpSDN, when a connection is established, \gls{TCP} waits to fill the \gls{MTU} before transmitting received messages from application. 
% Most TCP implementations use {Nagle's Algorithm} \cite{nagle1984congestion}, which means for messages less than \gls{MSS} size, the protocol waits to receive more bytes from an application before sending a packet.
% Although this method reduces message transmission overhead, it also results in longer message delivery delay.
% In the presence of packet loss, this method can affect message delivery delay.
% On the other hand, there is no such algorithm to slow down message transmission.

% Every payload received from application is queued on the streams and \gls{QUIC} multiplexing mechanism picks the streams payload, packs them in a \gls{QUIC} packet and transmits them.

\section{Related Work}
\label{relatedWork}
% The effect of transport layer protocols on the efficiency of southbound \gls{SDN} protocols has not been explored.
In this section, we first review the widely-used methods to enhance the scalability of \glspl{SDN} and then review the existing studies on the performance and applications of \gls{QUIC} protocol.

\subsection{SDN Scalability}
The communication overhead and delay between a controller and its associated switches have been explored in the literature.
The primary methods used to mitigate these overheads are: 
(i) increasing each switch's autonomy to handle flows, 
(ii) selecting optimal controller placement, 
and 
(iii) using multiple controllers to reduce switch-controller distances. 

To reduce the amount of switch-controller communication, Hedera \cite{al2010hedera} allows switches to handle mice flows via \gls{ECMP}, and switches only consult the controller when dealing with elephant flows. 
DIFANE \cite{yu2010scalable} distributes OpenFlow wildcards across switches to allow them to perform local routing.
Curtis et al. \cite{curtis2011devoflow} show that polling statistical data from switches reduces flow rule setup rate.
They also demonstrate that the low bandwidth between a switching appliance's CPU and ASIC introduces a significant communication delay between switch and controller when installing new flow rules.
They propose DevoFlow, which devolves the control of many flows back to switches, and the controller only targets significant flows.
DevoFlow uses wildcard rules to reduce the number of interactions with the controller, while also reducing the usage of \gls{TCAM}.
Mahout \cite{curtis2011mahout} uses sender's \gls{TCP} buffer size to identify mice flows and decide whether communication with the controller is necessary or not.
Kim et al. \cite{kim2014flow} propose a flow management scheme to reduce the number of OpenFlow \texttt{Packet\_in} messages sent to the controller, thereby reducing the network overhead caused by entry misses in a flow table.  
Their proposed scheme reduces table misses by maintaining inactive flow entries for as long as possible.
The inactive flow entries are maintained as long as the flow table has space; inactive flow entries are deleted once the flow table starts filling up.
Qin et al. \cite{qin2018SDNController} analyzed controller-switch and inter-controller traffic overheads in networks with varying numbers of nodes.
They show that the relationship between the amount of control traffic and the number of nodes in a network is linear. 
They model and propose a solution to the controller placement problem, reducing device management delay by 25\%.
Van Bemten et al. \cite{van2019empirical} use switches from multiple vendors and demonstrated that switch management operations are not predictable and reliable.
For example, with Pica switches, as the number of \texttt{Flow\_mod} messages per second increase, the switch shows two behaviors: the number of ignored rules increases, and some rules are reported to be installed while they have not been.

Onix \cite{koponen2010onix} provides a wide range of primitives for developing control applications in environments such as \gls{WAN} and public clouds.
To simplify this process while maintaining scalability, \glspl{API} are provided for a  distributed implementation.
For example, control applications can utilize these \glspl{API} to access the information maintained by Onix instances. 
HyperFlow \cite{tootoonchian2010hyperflow} synchronizes the status of distributed controllers and provides control applications with uniform, consistent access to the overall network data.
Kandoo \cite{hassas2012kandoo} assumes local processing is available close to the switches.
Applications that rely on local information are assigned to the local controllers, while non-local applications run in a root controller.
Bera et al. \cite{bera2020dynamic} propose a dynamic controller assignment scheme to maximize controller reactivity in heterogeneous networks.  
They accomplish this by selecting a controller to manage new flows that arrive at switches in the network, such that controller-switch delay and protocol overheads are optimized.  
Disco \cite{phemius2014disco} targets synchronization among controllers that manage multiple, heterogeneous networks.
They use \gls{AMQP}, which utilizes \gls{TCP}, to support east-west communication among controllers; \gls{AMQP} allows controllers to subscribe and publish to topics.

Zhang et al. \cite{zhang2011resilience} propose a min-cut algorithm for controller placement to enhance communication reliability with controllers.
The network is first partitioned with the minimum inter-partition cut, and inside each partition, the node with minimum distance to other nodes is found.
Survivor \cite{muller2014survivor} uses path diversity as a metric of their formulated linear programming model to determine controller location.
Simulation results show that the connectivity loss of the proposed method is between 2 to 3x less than \cite{zhang2011resilience}.
Beheshti et al. \cite{beheshti2012fast} argue the importance of providing switches with alternative paths to connect to the controller as soon as the primary path is dropped.
The proposed routing algorithm takes into account both distance and resilience to path failures.

Despite the valuable insights these works provide into the scalability of \glspl{SDN}, the effects of transport layer protocols on southbound communication have not been studied.
The contributions of this paper are orthogonal to the existing works and can be leveraged to enhance \gls{SDN} scalability further using methods such as dynamic switch-to-controller assignment and predictive flow installation.

\subsection{QUIC Protocol}
Started as an experimental protocol in Google, QUIC gained traction due to its benefits over \gls{TCP} in several types of networks \cite{kumar2019implementation, wang2018performance, lychev2015secure}.
Existing studies present the performance benefits of \gls{QUIC} over \gls{TCP} when used for transferring messages of application protocols such as HTTP/1.1, HTTP/2, and HTTP/3.
Zheng et al.
\cite{zheng2018performance} compared \gls{QUIC} with \gls{TCP}/\gls{TLS} in HTTP/2 and show that \gls{QUIC} performs better in congested networks where the retransmission rate is high.
QUIC achieves a higher performance due to its packet pacing method, which reduces traffic burstiness and packet loss.
Also, \gls{QUIC} multiplexes several streams carrying data over one connection, which enables \gls{QUIC} to carry more data than \gls{TCP} during a single round trip.
Multiplexing is specifically beneficial where medium access is expensive, such as on wireless networks.
Biswal et al. \cite{biswal2016does} compared HTTP/3's (\gls{QUIC}) and HTTP/2's (TCP) page load times for large and small objects and showed that for 0\% loss, HTTP/2 (TCP) performs better than HTTP/3 (\gls{QUIC}). 
This is because with 0\% loss, there is no congestion and the traffic flow is adjusted by flow control.
Since \gls{QUIC} employs flow control on each stream and the overall connection, this management of flow control per stream introduces latency.
On the other side, in realistic scenarios where packet loss occurs, HTTP/3 (QUIC) outperforms HTTP/2 (TCP).
This is because HTTP/2 is affected by the \gls{HOL} blocking problem.
Das et al. \cite{das2014evaluation} compared HTTP/2 (SPDY), HTTP/1.1 (TCP) and HTTP/3 (QUIC) under different network parameters including page load time and packet loss.
Their experiments were run for Amazon Alexa's top 500 websites \cite{alexa500}.
Their results show that HTTP/1.1  outperforms HTTP3 and HTTP2 for very minute bandwidth (0.2 Mbps).
For a small bandwidth, the bandwidth-delay product ($bandwidth \times \gls{RTT}$) is small, which means the amount of data carried in a round trip is small too.
%Small bandwidth-delay product results in several round trips between end points for transferring data.
In HTTP/1.1, this translates into opening several \gls{TCP} connections, while in HTTP/2 or HTTP/3, this causes opening several streams in one transport connection.
Since HTTP/1.1 does not employ multiplexing, only the \gls{TCP} connections experiencing packet loss are affected.
On the other hand, HTTP/2 and HTTP/3 rely on a single transport connection with multiple streams; therefore, a single stream suffering from packet loss can affect the entire connection.
When the bandwidth is increased to 1 Mbps, the bandwidth-delay product increases, packet loss rate drops, and HTTP/3 outperforms HTTP1.1.

\section{Conclusion}
\label{conclusion}
As the need for the higher rate, lower-overhead, and faster communication between the data plane and control plane in \glspl{SDN} increases, the role of the transport protocol used by southbound protocols increases too.
In this paper, we studied the shortcomings of using \gls{TCP} and justified the benefits of \gls{QUIC} over TCP in \glspl{SDN}.
We presented the design and implementation of quicSDN, a novel architecture that enables the communication of the control plane and data plane over the QUIC protocol.
We presented the benefits of quicSDN over tcpSDN via analytical studies and empirical evaluations.

Some future work directions to enhance the quicSDN architecture are as follows.
% is implementing a kernel bypass method for QUIC-SDN communication. 
First, the proposed quicSDN architecture can be used to improve the efficiency of switch-to-controller assignment methods for purposes such as load balancing.
Second, the proposed architecture can be leveraged for faster, more dynamic communication between switches and controllers to perform predictive configurations.
Third, in the current implementation, we use \glspl{UDS} for inter-process communication, which involves the Linux kernel's network stack for processing the exchanged messages.
The use and study of alternative methods are left as future work.
For example, on end-devices such as servers running containers and \glspl{VM}, the proposed architecture for switches may be integrated with \gls{DPDK} to reduce inter-process communication overhead.
Fourth, throughput and latency can be further improved by bringing Ethernet, IP, and \gls{UDP} packet processing into userspace for \gls{QUIC} packets.
Fifth, the benefits of \gls{QUIC}'s advanced flow and congestion control mechanisms have yet to be fully analyzed and quantified in \glspl{SDN}.

% Another potential area for future work is enhancing MTU flexibility. 
% At present \gls{QUIC} does not support packet fragmentation.
% The maximum size of the payload in a \gls{QUIC} packet can be 1350 bytes, and the \textit{Don't Fragment (DF)} bit must be set to prevent fragmentation on IP layer.
%Currently, \gls{QUIC} packets have a maximum size of 1392 bytes for handshake packets \cite{shade2014quic}, which includes the Ethernet, IP, and \gls{UDP} headers. 
%Due to the fixed maximum size MTU in handshake packets, \gls{QUIC} does not allow for packet fragmentation \cite{quicFrag}. 

\section*{Acknowledgment}
This work has been partially supported by a gift fund from Arista Networks.
Also, the authors would like to thank Arista Networks for donating the equipment used for this research.

\ifCLASSOPTIONcaptionsoff
  \newpage
\fi

\bibliographystyle{IEEEtran}
\bibliography{references}


%\vfill


\end{document}

%% file: list_abbreviations.tex
\newcommand{\ovsdbserver}{{\fontfamily{lmss}\selectfont ovsdb-server}}

\newcommand{\quicclient}{{\fontfamily{lmss}\selectfont quic-client}}

\newcommand{\quicserver}{{\fontfamily{lmss}\selectfont quic-server}}

\newcommand{\connecting}{{\fontfamily{lmss}\selectfont CONNECTING}}
\newcommand{\ovsswitchd}{{\fontfamily{lmss}\selectfont ovs-switchd}}
\newcommand{\ngtcp}{{\fontfamily{lmss}\selectfont ngtcp$2$}}

\newcommand{\openflowryuapp}{{\fontfamily{lmss}\selectfont ryu-of}}
\newcommand{\ovsdbryuapp}{{\fontfamily{lmss}\selectfont ryu-ovsdb}}

\newcommand{\activeState}{{\fontfamily{lmss}\selectfont ACTIVE}}

\newcommand{\ryuovsdb}{{\fontfamily{lmss}\selectfont ryu-ovsdb}}

\newcommand{\ryuopenflow}{{\fontfamily{lmss}\selectfont ryu-of}}

\newacronym{OS}{OS}{Operating System}
\newacronym{NACK}{NACK}{Negative Acknowledgement}

\newacronym{SDK}{SDK}{Software Development Kit}

\newacronym{RPC}{RPC}{Remote Procedure Calls}

\newacronym{EOS}{EOS}{Extensible Operating System }

\newacronym{gNMI}{gNMI}{gRPC Network Management Interface}

\newacronym{gNOI}{gNOI}{gRPC Network Operations Interface}

\newacronym{gRIBI}{gRIBI}{gRPC Interface to a Network Element RIB}

\newacronym{QSH}{QSH}{QUIC Short Header}
\newacronym{QFH}{QFH}{QUIC Frame Header}

\newacronym{AEAD}{AEAD}{Authenticated Encryption with Associated Data}

\newacronym{RSSI}{RSSI}{Received Signal Strength Indicator}

\newacronym{DNS}{DNS}{Domain Name System}
\newacronym{RTT}{RTT}{Round Trip Time}
\newacronym{RTO}{RTO}{Retransmission Timeout}

\newacronym{IPC}{IPC}{Inter-Process Communication}
\newacronym{TCP}{TCP}{Transmission Control Protocol}
\newacronym{UDP}{UDP}{User Datagram Protocol}
\newacronym{UDS}{UDS}{Unix Domain Socket}

\newacronym{HOL}{HOL}{Head-of-Line}

\newacronym{WEP}{WEP}{Wired Equivalent Privacy }
\newacronym{ryu-ovsdb}{ryu-ovsdb}{RYU-Controller}
\newacronym{quic-client}{quic-client}{QUIC Client}
\newacronym{WPA}{WPA}{Wireless Protected Access}
\newacronym{PMK}{PMK}{Pairwise Master Key}
\newacronym{PTK}{PTK}{Pairwise Transition Keys}
\newacronym{GTK}{GTK}{Group Temporal Key}
\newacronym{EAP}{EAP}{Extensible Authentication Protocol}
\newacronym{RRM}{RRM}{Radio Resource Management}
\newacronym{BSSTM}{BSSTM}{BSS Transition Management}
\newacronym{BSS}{BSS}{Basic Service Set}
\newacronym{SSID}{SSID}{Service Set Identifier}
\newacronym{BSSID}{BSSID}{BSS Identifier}
\newacronym{CSA}{CSA}{Channel Switching Announcement}
\newacronym{EDCF}{EDCF}{Enhanced Distributed Coordination Function}
\newacronym{AIFS}{AIFS}{Arbitration Inter-Frame Space}

\newacronym{PSM}{PSM}{Power Save Mode}
\newacronym{PTO}{PTO}{Probe Timeout}
\newacronym{APSD}{APSD}{Adaptive Power Save Delivery}

\newacronym{DBS}{DBS}{Disassociation-Based Steering}

\newacronym{RTP}{RTP}{Real-time Transport Protocol}

\newacronym{CPU}{CPU}{Central Processing Unit}

\newacronym{SNMP}{SNMP}{Simple Network Management Protocol}
\newacronym{AMQP}{AMQP}{Advanced Message Queuing Protocol}

\newacronym{TLS}{TLS}{Transport Layer Security}

\newacronym{DTLS}{DTLS}{Datagram Transport Layer Security}

\newacronym{QUIC}{QUIC}{Quick UDP Internet Connection}

\newglossaryentry{NOS}
{
  name={NOS},
  description={Network Operating System},
  first={\glsentrydesc{NOS} (\glsentrytext{NOS})},
  plural={NOSs},
  firstplural={\glsentrydesc{NOS}s (\glsentryplural{NOS})}
}

\newglossaryentry{VM}
{
  name={VM},
  description={Virtual Machine},
  first={\glsentrydesc{VM} (\glsentrytext{VM})},
  plural={VMs},
  firstplural={\glsentrydesc{VM}s (\glsentryplural{VM})}
}

\newglossaryentry{WAN}
{
  name={WAN},
  description={Wide Area Network},
  first={\glsentrydesc{WAN} (\glsentrytext{WAN})},
  plural={WANs},
  firstplural={\glsentrydesc{WAN}s (\glsentryplural{WAN})}
}

\newglossaryentry{FD}
{
  name={FD},
  description={File Descriptor},
  first={\glsentrydesc{FD} (\glsentrytext{FD})},
  plural={FDs},
  firstplural={\glsentrydesc{FD}s (\glsentryplural{FD})}
}

\newglossaryentry{API}
{
  name={API},
  description={Application Programming Interface},
  first={\glsentrydesc{API} (\glsentrytext{API})},
  plural={APIs},
  firstplural={\glsentrydesc{API}s (\glsentryplural{API})}
}

\newglossaryentry{SDNing}
{
  name={SDN},
  description={Software-Defined Networking},
  first={\glsentrydesc{SDN} (\glsentrytext{SDN})},
  plural={SDNs},
  firstplural={\glsentrydesc{SDN}s (\glsentryplural{SDN})}
}

\newglossaryentry{SDN}
{
  name={SDN},
  description={Software-Defined Network},
  first={\glsentrydesc{SDN} (\glsentrytext{SDN})},
  plural={SDNs},
  firstplural={\glsentrydesc{SDN}s (\glsentryplural{SDN})}
}

\newacronym{BESS}{BESS}{Berkeley Extensible Software Switch}

\newacronym{CLI}{CLI}{Command Line Interface}

\newacronym{ECMP}{ECMP}{Equal-Cost Multi-Path}

\newacronym{RED}{RED}{Random Early Detection}

\newacronym{ODL}{ODL}{OpenDayLight}
\newacronym{OVS}{OVS}{Open vSwitch}
\newacronym{OVSDB}{OVSDB}{Open vSwitch Database}
\newacronym{VNF}{VNF}{Virtual Network Function}
\newacronym{NNF}{NNF}{Native Network Function}

\newacronym{PCP}{PCP}{Priority Code Point}

\newacronym{NAT}{NAT}{Network Address Translation}

\newacronym{NUMA}{NUMA}{Non-Uniform Memory Access}

\newacronym{CPE}{CPE}{Customer Premise Equipment}

\newacronym{DSCP}{DSCP}{Differentiated Services Code Point}

\newacronym{vSG}{vSG}{Virtual Subscriber Gateway}

\newacronym{RGW}{RGW}{Residential Gateway}
\newacronym{vRGW}{vRGW}{Virtual RGW}

\newacronym{VF}{VF}{Virtual Function}
\newacronym{PF}{PF}{Physical Function}
\newacronym{ACK}{ACK}{Acknowledgement}
\newacronym{SACK}{SACK}{Selective Acknowledgement}

\newacronym{ARP}{ARP}{Address Resolution Protocol}

\newacronym{DTIM}{DTIM}{Delivery Traffic Indication Message}
\newacronym{AP}{AP}{Access Point}
\newacronym{TWT}{TWT}{Target Wake-up Time}
\newacronym{OPS}{OPS}{Opportunistic Power Save}
\newacronym{PCF}{PCF}{Point Coordination Function}
\newacronym{PIFS}{PIFS}{PCP Inter Frame Space}
\newacronym{DIFS}{DIFS}{Distributed Inter Frame Space}
\newacronym{BSR}{BSR}{Buffer Status Report}
\newacronym{OCW}{OCW}{OFDM Contention Window}

\newacronym{QTP}{QTP}{Quite Time Period}
\newacronym{AU}{AU}{Airtime Utilization}

\newacronym{LVAP}{LVAP}{Light Virtual AP}
\newacronym{EDCA}{EDCA}{Enhanced Distributed Channel Access}

\newacronym{HCCA}{HCCA}{Hybrid Controlled Channel Access}

\newacronym{DC}{DC}{Datacenter}
\newacronym{CO}{CO}{Central Office}

\newacronym{CORD}{CORD}{Central Office Re-architected as a Datacenter}
\newacronym{MEC}{MEC}{Mobile Edge Computing}
\newacronym{ASP}{ASP}{Application Service Provider}
\newacronym{ES}{ES}{Edge Service}
\newacronym{NSP}{NSP}{Network Service Provider}

\newacronym{NV}{NV}{Network Virtualization}
\newacronym{NFV}{NFV}{Network Function Virtualization}
\newacronym{NFVI}{NFV}{NFV Infrastructure}
\newacronym{MANO}{MANO}{Management and Orchestration}

\newacronym{vNIC}{vNIC}{virtual NIC}
\newacronym{pNIC}{pNIC}{physical NIC}

\newacronym{TCAM}{TCAM}{Ternary Content Addressable Memory}
\newacronym{RSS}{RSS}{Receive Side Scaling}

\newacronym{DPDK}{DPDK}{Data Plane Development Kit}
\newacronym{UIO}{UIO}{User-space I/O}

\newacronym{EMC}{EMC}{Exact Match Cache}

\newacronym{SR-IOV}{SR-IOV}{Single Root-Input/Output Scaling}
\newacronym{PMD}{PMD}{Poll Mode Driver}

\newacronym{TSS}{TSS}{Tuple Space Search}
\newacronym{dpcls}{dpcls}{data path classifier}

\newacronym{FDK}{FDK}{Fog Development Kit}

\newacronym{TSN}{TSN}{Time Sensitive Networking}
\newacronym{AFDX}{AFDX}{avionics full-duplex switched Ethernet}
\newacronym{CAN}{CAN}{Controller Area Network}

\newacronym{OCI}{OCI}{Open Carrier Interface }

\newacronym{CBR}{CBR}{constant bit rate}

\newacronym{LSTM}{LSTM}{Long Short-Term Memory}

\newacronym{TSO}{TSO}{TCP Segmentation Offload}
\newacronym{GRO}{GRO}{Generic Receive Offload}

\newacronym{NIC}{NIC}{Network Interface Card}
\newacronym{MTU}{MTU}{Maximum Transmission Unit}
\newacronym{MSS}{MSS}{Maximum Segment Size}

\newacronym{QoS}{QoS}{Quality of Service}

\newacronym{HTB}{HTB}{Hierarchical Token Bucket}
\newacronym{trTCM}{trTCM}{Two-Rate, Three Color Marker}

\newacronym{PDF}{PDF}{probability density function}